\documentclass[%
 aip,
 rsi,
 amsmath,amssymb,
 reprint,%
]{revtex4-1}

\usepackage{graphicx}
\usepackage{dcolumn}
\usepackage{bm}

\usepackage[utf8]{inputenc}
\usepackage[T1]{fontenc}
\usepackage{mathptmx}
\usepackage{etoolbox}
\draft 

\begin{document}


\title{The chemRIXS Instrument for the LCLS-II X-Ray Free Electron Laser} 



\author{David J. Hoffman}
\author{Douglas Garratt}
\author{Matthew Bain}
\author{Christina Y. Hampton}
\affiliation{SLAC National Accelerator Laboratory, Menlo Park, CA, USA}

\author{Benjamin I. Poulter}
\affiliation{Stanford PULSE Institute, SLAC National Accelerator Laboratory, Menlo Park, CA, USA.}

\author{Jyoti Joshi}
\author{Giacomo Coslovich}
\author{Frank P. O'Dowd}
\author{Daniel P. DePonte}
\author{Alexander H. Reid}
\author{Lingjia Shen}
\author{Daniel Jost}
\author{Mina R. Bionta}
\author{Joshua J. Turner}
\author{Ming-Fu Lin}
\author{Philip Heimann}
\author{Stefan P. Moeller}
\author{Jake D. Koralek}

\author{Tyler Johnson}
\author{K Ninh}
\author{Raybel Almeida}
\author{Adam Berges}
\author{Stephanie Fung}
\affiliation{SLAC National Accelerator Laboratory, Menlo Park, CA, USA}

\author{Shuai Li}
\author{Eetu Pelimanni}
\affiliation{Chemical Sciences and Engineering Division, Argonne National Laboratory, Lemont, IL, USA.}

\author{Amke Nimmrich}
\author{Munira Khalil}
\affiliation{Department of Chemistry, University of Washington, Seattle, WA, USA.}

\author{Linda Young}
\affiliation{Chemical Sciences and Engineering Division, Argonne National Laboratory, Lemont, IL, USA.}
\affiliation{Department of Physics and James Franck Institute, The University of Chicago, Chicago, IL, USA.}

\author{Thomas J.A. Wolf}
\affiliation{SLAC National Accelerator Laboratory, Menlo Park, CA, USA}
\affiliation{Stanford PULSE Institute, SLAC National Accelerator Laboratory, Menlo Park, CA, USA.}

\author{Kristjan Kunnus}
\email[]{kristjan@slac.stanford.edu}
\author{Georgi L. Dakovski}
\email[]{dakovski@slac.stanford.edu}
\affiliation{SLAC National Accelerator Laboratory, Menlo Park, CA, USA}

\date{\today}

\begin{abstract}
The chemRIXS instrument at the Linac Coherent Light Source offers new opportunities for studying solution-phase systems with time-resolved soft X-ray spectroscopy through the recently commissioned high-repetition-rate LCLS-II X-ray free electron laser. The orders-of-magnitude X-ray flux improvement provided by the superconducting accelerator, combined with corresponding advances in the optical laser system and the liquid jet recirculation system, enables studies on dilute systems with high signal-to-noise compared to what was possible with the LCLS-I copper accelerator. These capabilities open up time-resolved X-ray absorption spectroscopy and resonant inelastic X-ray scattering to entirely new classes of samples, as well as enabling the development of new soft X-ray spectroscopies on liquid samples. An overview of the beamline components and the first LCLS-II commissioning results are presented.

\end{abstract}


\maketitle 


\section{Introduction}

\begin{figure*}
    \centering
    \includegraphics[width=1\linewidth]{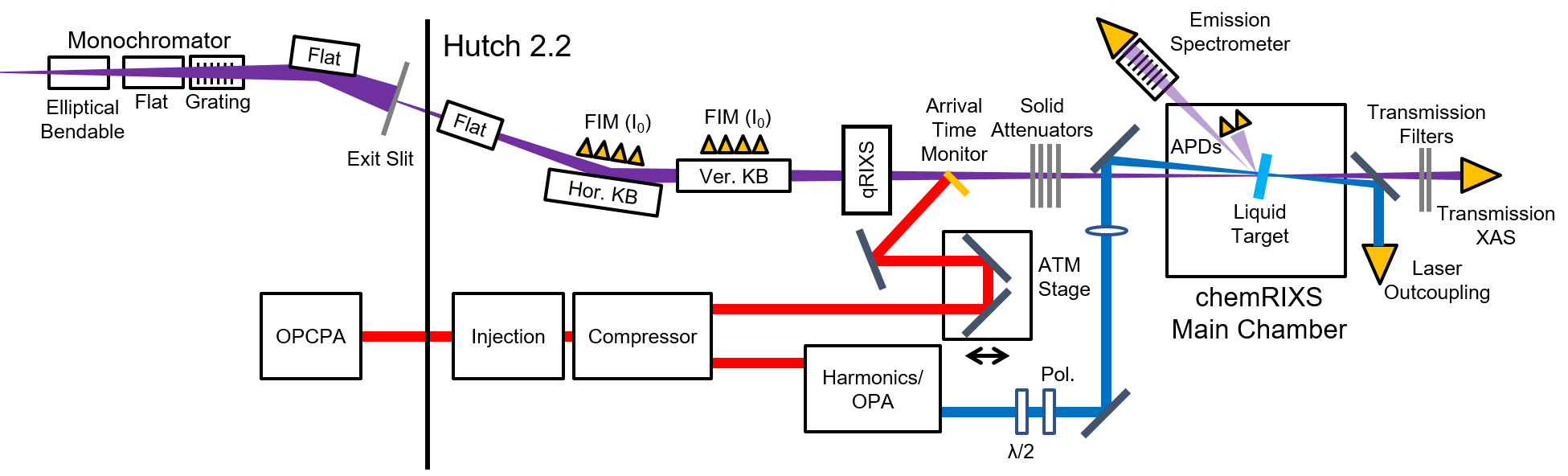}
    \caption{Schematic overview of the principal X-ray beamline components and optical laser components for the chemRIXS instrument. Purple indicates the X-ray beam, red indicates the OPCPA fundamental (800 nm), and blue indicates the variable-wavelength optical laser beam delivered to the chemRIXS IP. KB: Kirkpatrick-Baez Mirror, ATM: Arrival Time Monitor, OPCPA: Optical Parametric Chirped Pulse Amplifier, OPA: Optical Parametric Amplifier, FIM: Fluorescence Intensity Monitor, APDs: Avalanche Photodiodes, Pol.: Polarizer, $\lambda/2$: half-waveplate.}
    \label{fig:BL}
\end{figure*}

The chemRIXS Instrument at the Linac Coherent Light Source (LCLS) at the SLAC National Accelerator Lab is a successor instrument to the liquid jet endstation\cite{kunnus2012rosi} used at the Soft X-ray Research (SXR) instrument formerly at LCLS.\cite{schlotter2012rsi,dakovski2015jsr} ChemRIXS is dedicated to ultrafast soft X-ray spectroscopy on solution-phase systems, which provides great opportunities for studying chemical systems, as the soft X-ray transitions (250-1600 eV) involve core electrons in light elements and third-row transition metals.\cite{worner2025nrc,jay2022arpc} The energies of these transitions have element-dependent binding energies that are usually separated by 10s or 100s of eVs, making the techniques element selective. Combined with femtosecond optical lasers and X-ray free electron lasers (XFELs), which provide tunable, high intensity, femtosecond X-ray pulses,\cite{emma2010np,bergman2017x} these methods can be used for tracking the evolution of both unoccupied and occupied orbitals during photochemical reactions on ultrafast timescales.

X-ray absorption spectroscopy (XAS) probes resonances between core to valence electronic orbitals,\cite{worner2025nrc} which can then provide details on the overlap between specific atomic sites (through these core electronic orbitals) and a given valence orbital. After photoexcitation by an optical laser, the time-resolved XAS spectra can report on the evolution of the solvation environment, the depleted ground state, and the evolution of excited states in relation to these core electron orbitals. Complementary information can be obtained by examining the X-ray emission channels. Resonant inelastic X-ray scattering (RIXS)\cite{jay2022arpc,wernet2015n,degroot2024nrmp,schlappa2025jsr} is used to track various energy transfer pathways within the system of interest by spectrally resolving emitted X-ray photons. For chemical systems, this can be used to examine the couplings between occupied and unoccupied valence orbitals with the same element and site specificity provided by XAS. In a single-electron picture of the RIXS process, the monochromatic X-rays first trigger a resonant transition from a core electron to an unoccupied valence orbital. Occasionally, an electron from a different orbital will drop down into the core hole, which emits a photon at a different energy. Time-resolved RIXS can then provide information on how these occupied orbitals evolve after laser excitation, and how they overlap with the corresponding unoccupied orbitals probed by time-resolved XAS.

However, there are substantial technical restrictions that have slowed the development of the field: the contrast between solute and solvent is often poor, the fluorescence yields are low due to the high probability of instead emitting electrons through Auger-Meitner decay, and the efficiency of soft X-ray grazing-incidence grating spectrometers is low.\cite{jay2022arpc} For these reasons, time-resolved RIXS in particular is extremely photon-hungry, which, when combined with the additional inherent difficulties of measuring liquid samples in high vacuum, has limited its application to only a handful of high-concentration chemical systems in measurements at previous-generation XFELs.\cite{wernet2015n,jay2018jpcl,kjellsson2020prl}

ChemRIXS addresses these issues by leveraging the orders-of-magnitude improvement in repetition rate offered by the recently commissioned LCLS-II free electron laser at SLAC to use these X-ray photon-hungry experimental techniques. ChemRIXS then uses a corresponding suite of detectors, liquid jet delivery system, and high-repetition-rate optical laser system to enable time-resolved XAS, RIXS, and developing nonlinear soft X-ray spectroscopies on solution-phase systems. In this article, an overview of chemRIXS and its associated beamline, laser infrastructure, detectors, and sample delivery system is presented. We conclude with an overview of early scientific accomplishments made using LCLS-I and the first commissioning results from chemRIXS with the high-repetition-rate LCLS-II beam.

\section{Instrument Components}

The chemRIXS instrument is located at the second interaction point of Hutch 2.2 of the Near Experimental Hall of the LCLS. A schematic diagram of the principal beamline and instrument components can be found in Fig.~\ref{fig:BL}. A full description of the beamline optics and other components shared between the three interaction points (IPs; the materials science endstation qRIXS, chemRIXS, and future roll-up endstations respectively) can be found elsewhere,\cite{nicolas2022p,odowd2018pmesrei} but a brief overview is presented here.

\subsection{XFEL Source}

ChemRIXS utilizes the soft X-ray (SXR) line of LCLS and was designed around the high-flux capabilities of the LCLS-II accelerator, although for the first few years of its operation (2021-2022) it used the LCLS-I SXR beam. The main difference between the LCLS-I and LCLS-II beam is the repetition rate, where LCLS-I operates at 120 Hz while LCLS-II will eventually be able to operate at nearly 1 MHz. To date, the working repetition rate of LCLS-II has been at 33.2 kHz or at subharmonics thereof (16.6 kHz or 8.3 kHz).

The core X-ray and beamline parameters are summarized in Table~\ref{xray-table}. The photon energies produced from the variable gap SXR undulators span 200-1600 eV. In the standard self-amplified spontaneous emission (SASE) operation, the X-ray pulses have energies of hundreds of microjoules and pulse durations of tens of femtoseconds. Additional operating modes (self-seeding,\cite{cocco2013} attosecond pulses,\cite{duris2020np,franz2024np} two-color\cite{guo2024np}) that have been used at LCLS-I are also being developed for LCLS-II.

\begin{table}[!b]
\caption{Core X-ray and beamline parameters for monochromatic operation. Values at chemRIXS IP unless otherwise noted.}
\label{xray-table}
\begin{ruledtabular}
\begin{tabular}{ll}
 \textbf{Beamline Parameter} & \textbf{Typical Performance}\\
 \hline
 Energy Range (eV)       & 300-1600\\
 Repetition Rate (Hz)    & Up to 33163\footnote{LCLS-II, as of September 2025}; 120\footnote{LCLS-I}\\
 Pulse Energy ($\mu$J)   & 400\footnote{Before monochromator, optimum, at 400-500 eV} \\
 Focus ($\mu$m)          & 10-1000 \\
 Pulse Duration (fs)     & 30\footnote{Before monochromator} \\
 Transmission& 1-2\%\footnote{Monochromatized} \\
 X-ray Polarization            & Horizontal \\
 Resolving Power (E/$\Delta$E)        & 2000 \\
 Photon Flux at IP ($s^{-1}$)& Up to $2 \times 10^{15}$ $^\mathrm{a}$\\
 
\end{tabular}
\end{ruledtabular}
\end{table}

\subsection{Beamline Components}

Before reaching the beamline optics, the soft X-ray pulses produced by LCLS go through a gas monitor detector (GMD)\cite{tiedtke2014oea} and gas attenuator which provides both average and shot-by-shot pulse energy measurements as well as controllable attenuation of the X-ray beam (generally using either N$_2$ or Ar depending on the photon energy).

\begin{figure}[bt]
    \centering
    \includegraphics[width=1\linewidth]{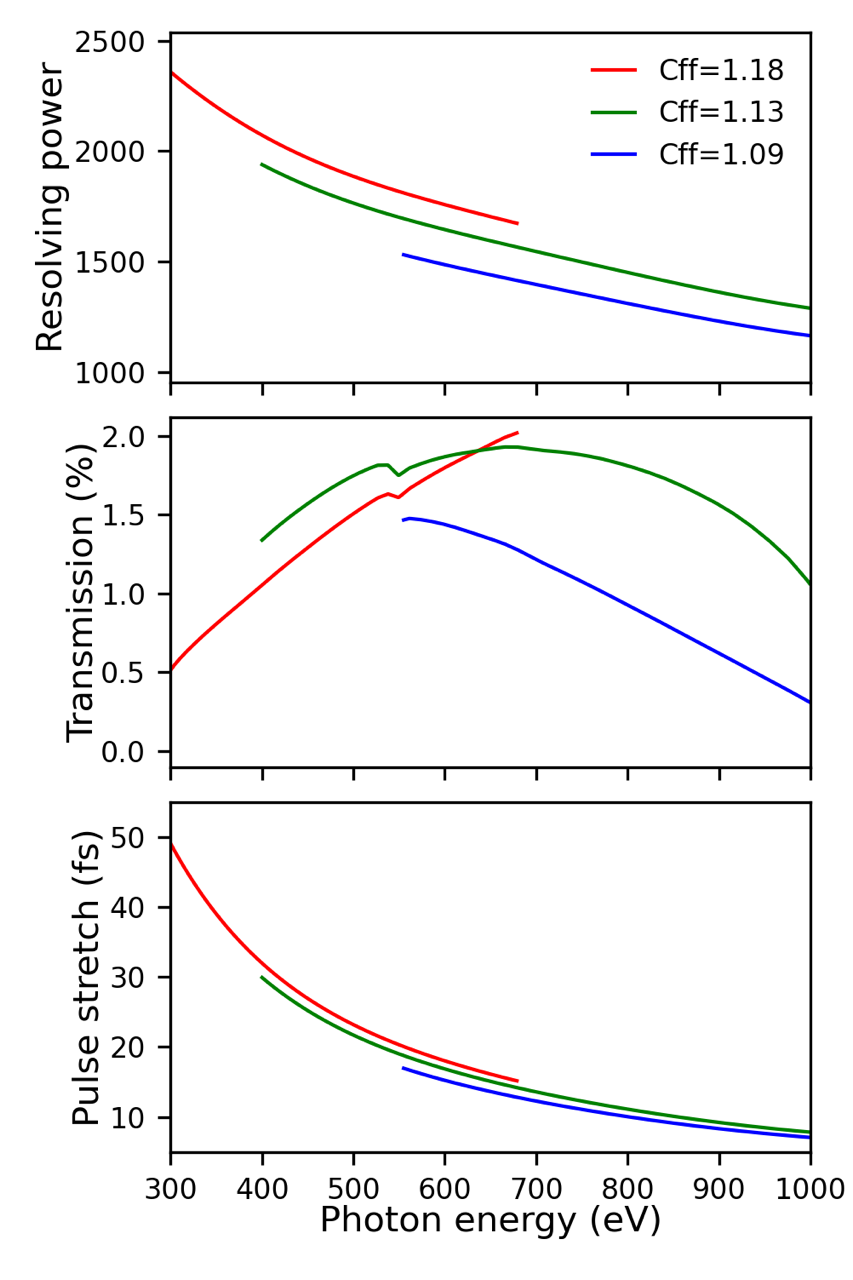}
    \caption{Monochromator resolving power ($\Delta E/E$), beamline transmission, and pulse stretching for the low resolution grating used at chemRIXS for 300 to 1000 eV for select constant focal factors (Cff) achievable by the monochromator. All values FWHM. Resolving power and transmission correspond to approx. 30 $\mu$m exit slit width  - equal to the width of monochromatic image at the exit slit plane. Transmission calculated for B$_4$C coating. More details can be found in a prior publication.\cite{nicolas2022p}}
    \label{fig:LRG}
\end{figure}

The pulses are next sent through a monochromator comprised of a vertically-focusing elliptical bendable mirror, a flat mirror, and a flat variable-line-spacing (VLS) blazed grating.\cite{nicolas2022p} Multiple gratings are mounted in the monochromator: chemRIXS almost exclusively uses the "low resolution grating." Its central line density of 50 lines/mm (resolving power \textasciitilde 2000, generally giving resolutions of several hundred meV) was chosen to provide a balance between energy and time resolution for the target chemRIXS science cases. The designed resolving power, beamline transmission, and pulse stretching are shown in Fig.~\ref{fig:LRG}.

The first-order energy-dispersed beam is focused at an exit slit approximately 20 m away, which then yields the monochromatized soft X-ray beam that is sent to the hutch. Alternatively to the monochromatized first-order beam, the zero-order “pink” beam can also be sent through the monochromator for nonlinear X-ray experiments discussed briefly below.

To steer and focus the beam, there are a pair of flat mirrors and a pair of bendable Kirkpatrick-Baez (KB) mirrors which allow the beam to be pointed and focused independently along the x- and y-axes. The KBs are designed to be able to focus the X-ray beam at any of the three interaction points on Beamline 2.2. When focused to the chemRIXS IP, the horizontal KB images the X-ray source in the undulator with approximately a factor of 16 demagnification (assuming 145 m distance to the source, source distance can vary) and the vertical KB images the exit slit with a demagnification of 3.2.   The typical focused spot size is \textasciitilde 10 $\mu$m FWHM at the chemRIXS IP and the beam can be defocused up to 1000 $\mu$m. The KB mirrors are also equipped with fluorescence intensity monitors (FIMs),\cite{heimann2019jsr,heimann2025jpcs} which are sets of four avalanche photodiodes (APDs) and microchannel plates (MCPs) that serve as normalization $I_0$ detectors for the monochromatized beam on a shot-by-shot basis, even at the eventual full LCLS-II repetition rate of 929 kHz.

A set of four solid attenuator paddles equipped with Al and Co filters of various thicknesses are just upstream of the main chemRIXS chamber. These filters allow for the precise attenuation of the X-ray beam without reducing the signal on the upstream FIM $I_0$ detectors or on the arrival time monitor discussed below, and can help attenuate unwanted harmonics from the XFEL.

Due to the large number of mirrors involved in the beamline, most of the X-ray optics are equipped with two separate coatings to cover the entire soft X-ray region accessible at LCLS (\textasciitilde 200-1600 eV): a boron carbide (B$_4$C) coating for photon energies \textless 1000 eV and a rhodium coating for energies \textgreater 1000 eV. The two coatings can be switched by translating the optics perpendicular to the beam. Even with the targeted coatings, typical beamline transmissions in monochromatic operation are about 1\% for photon energies \textless 1000 eV and between 0.1-0.01\% for \textgreater 1000 eV.

For many experiments at chemRIXS it is desirable to scan the incoming photon energy. Typically, this is done by jointly scanning the gap of the undulator magnets (thus changing the energy of the peak of the SASE spectrum) with the monochromator optics’ pitches. This configuration allows energy scans spanning tens of eV, which is much greater than the SASE pulse bandwidth (typically \textasciitilde 0.5\% of the photon energy). 

\subsection{The chemRIXS Chamber}

\begin{figure*}
    \centering
    \includegraphics[width=0.7\linewidth]{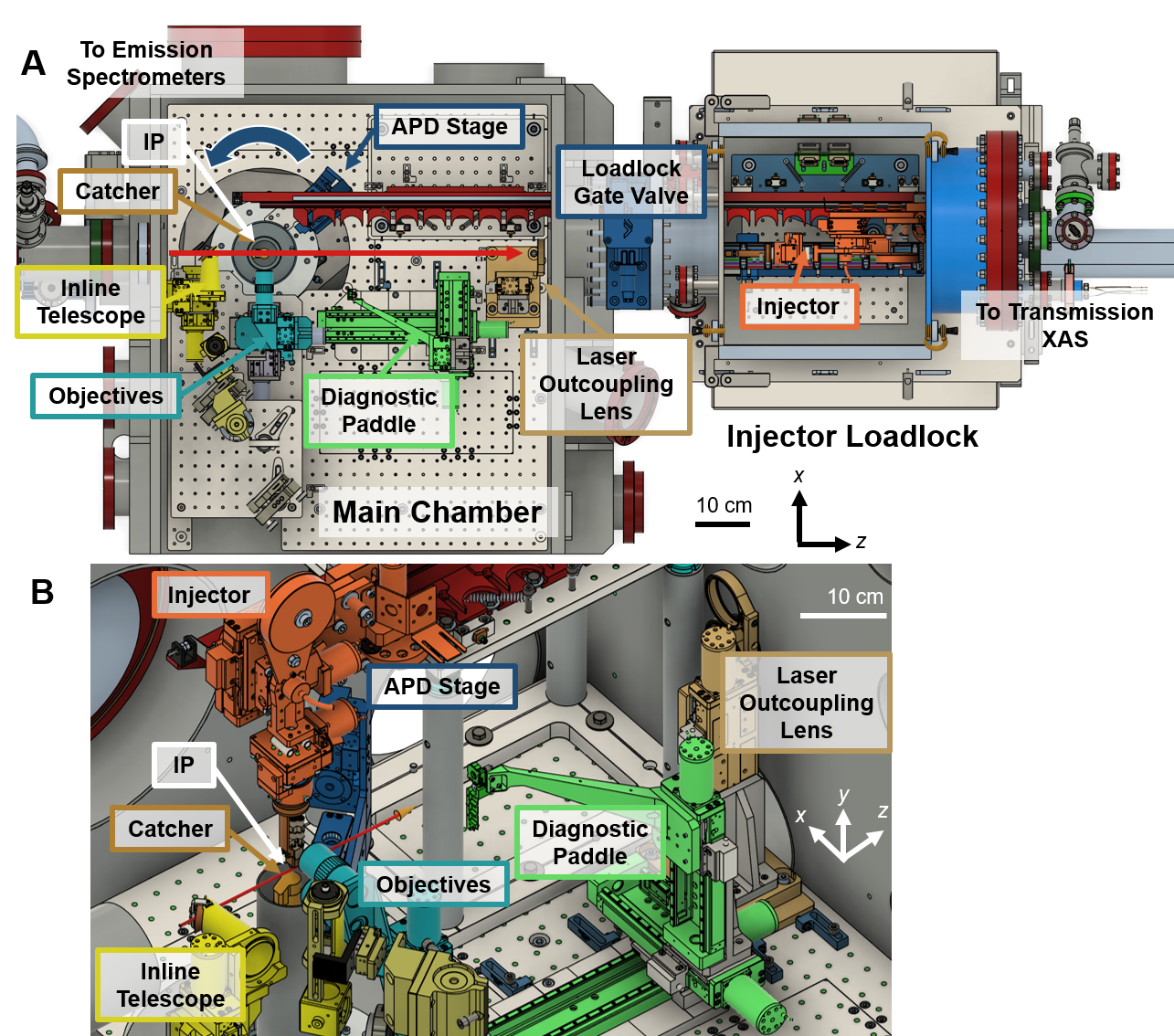} 
    \caption{A. Top down view of the main chemRIXS chamber and injector load-lock chamber. The X-ray path is indicated with a red arrow. The principle components are indicated. B. Isometric view of the chemRIXS IP. The injector is shown inserted in its experimental configuration.}
    \label{fig:chemrixs}
\end{figure*}

A CAD model illustrating the major components of the chemRIXS chamber is shown in Fig.~\ref{fig:chemrixs}, with the path of the X-rays indicated with a red arrow. The main vacuum chamber has dimensions of $0.85 \times 0.75 \times 0.7$ m and is connected to a secondary injector loadlock chamber (dimensions $0.43 \times 0.39 \times 0.39$ m) with a gate valve. The injector loadlock chamber is designed to allow for the independent venting of the injector stage and liquid jet nozzles without having to vent the main chemRIXS chamber. This design allows damaged or clogged nozzles to be replaced in under an hour, which is a third of the time required to safely vent and pump down the full chemRIXS chamber.

The injector (highlighted orange in Fig.~\ref{fig:chemrixs}) is the 5-axis stage that holds the liquid jet nozzles (discussed in detail below). The injector stage translates almost a meter along a Geneva drive rack-and-pinion and rotates 90° along its x-axis when transferring between the loadlock (as shown in Fig.~\ref{fig:chemrixs}A) and the IP (as shown in Fig.~\ref{fig:chemrixs}B). The full insertion or retraction procedure takes approximately 5 minutes to complete. The injector stage can also move in the plane orthogonal to the beam with 10 $\mu$m step precision and rotate along the nozzle axis.

In the IP, the catcher (brown in Fig.~\ref{fig:chemrixs}) is aligned to the liquid jet and takes the liquid out of the vacuum chamber. It is then connected to the liquid recirculator system (described below). The catcher has a conical shape with a 400-500 $\mu$m diameter hole at the top and is connected to a 3-axis motorized stage to align the catcher to the liquid jet.

The liquid jet/catcher alignment is facilitated using two in-vacuum imaging systems. The first is an inline telescope (yellow in Fig.~\ref{fig:chemrixs}), which has a holey mirror (mirror with a circular cutout to allow the X-ray beam to pass) in the path of the X-ray and laser beams. Fiber-coupled LEDs and a beamsplitter are used to provide in-line illumination of the target. The collection mirror is aligned into a telescope (Questar FR-1 MKIII) before being resolved on a camera (Manta G-419B NIR). The collection mirror is on a tip/tilt and 3-axis translation stage to permit remote alignment in vacuum.

The second imaging system (cyan in Fig.~\ref{fig:chemrixs}) uses infinity corrected long-working-distance microscope objectives ($7.5\times$ / 0.21 NA Mitutoyo M Plan Apo SL or $20\times$ / 0.28 NA Mitutoyo M Plan Apo SL) with bright field illumination from another fiber-coupled LED. The objectives are on a 3-axis stage which allows for remote alignment and toggling between the two objectives in vacuum. The objective's light path is then directed into a 200 mm tube lens and imaged on a camera (Manta G-419B NIR). Both imaging systems are also equipped with filter slider stages to insert laser light filters as needed.

The diagnostic paddle (green in Fig.~\ref{fig:chemrixs}) has a variety of targets that can be inserted into the beam. This includes YAG screens for characterizing the X-ray and laser spot sizes and establishing spatial overlap (using the inline telescope for imaging), a fast photodiode used for establishing coarse timing (of order 100s of ps) in the IP between the X-ray and laser pulses, and a set of solid reference targets (e.g., hexagonal boron nitride or cobalt oxide) used for calibrating the emission spectrometers. The YAG screen is also used for establishing fine timing with the laser by observing the onset of X-ray-induced reflectivity (using the laser outcoupling lens also seen in Fig.~\ref{fig:chemrixs}).

ChemRIXS can also mount solid targets to a four-axis stage ($X$, $Y$, $Z$, $\theta$) with a liquid nitrogen cryostat or a simple paddle for room temperature measurements. The manipulator is located above the interaction point, so it is not shown in Fig.~\ref{fig:chemrixs}.

\subsection{Detectors}

The chemRIXS endstation standard liquid configuration is equipped with three main detection channels which, in general, can all be used in parallel. Avalanche photodiodes (APD) are used to measure shot-by-shot total fluorescence yield (TFY) XAS. The integrating detectors include the emission spectrometer, which measures partial fluorescence yield (PFY-XAS) and resonant inelastic X-ray scattering (RIXS), and the transmission detector, which measures XAS by monitoring the transmission through the liquid target. The different XAS detection channels offer different advantages depending on the experiment: the transmission XAS is useful in strongly absorbing samples and provides direct measurement of absorption cross-sections, the TFY-XAS provides the best time resolution and works best below the solvent absorption edge, and the PFY-XAS offers the best sensitivity just above the solvent absorption edge as it can filter out solvent fluorescence.

Beyond the standard configuration detectors, a transmission spectrometer is also being developed for use in nonlinear X-ray experiments with the pink X-ray beam, where it is desirable to resolve the X-ray spectrum after transmission through the liquid target.

\subsubsection{Emission Spectrometer}

\begin{figure}[bt]
    \centering
    \includegraphics[width=1\linewidth]{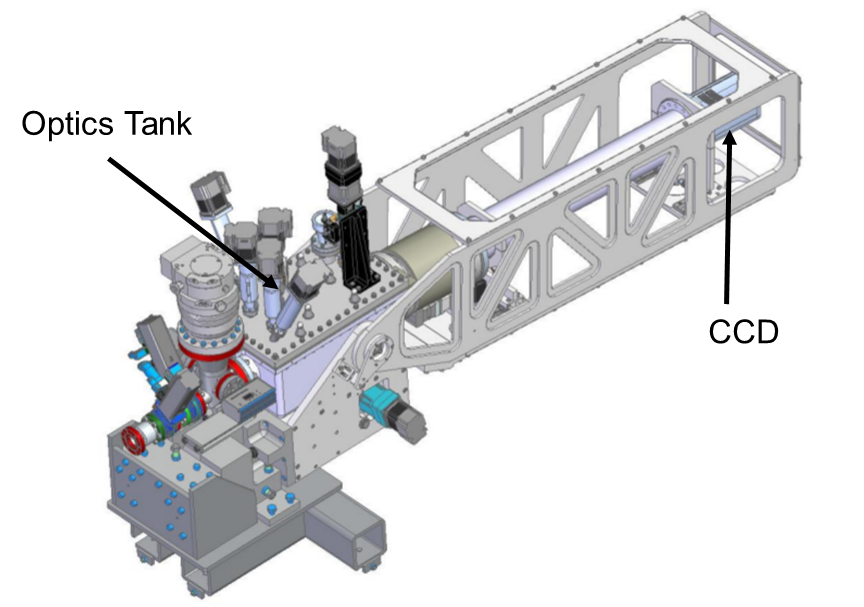}
    \caption{Model of the VLS spectrometer.}
    \label{fig:VLS}
\end{figure}

The chemRIXS endstation is currently equipped with a portable variable line spacing (VLS) grating emission spectrometer that is typically mounted at a 45° backscattering direction. The spectrometer uses the design from Chuang et al.\cite{chuang2017rosi} based on a Hettrick-Underwood optical scheme.\cite{hettrick1986aov2i2p4} A model of the VLS spectrometer can be seen in Fig.~\ref{fig:VLS} and full tables of parameters can be found in Table~\ref{vls-table}. It consists of a focusing mirror and a VLS grating (1200 lines/mm, resolving power \textasciitilde 1500) and a CCD camera (Andor Newton SO) with an overall efficiency of \textasciitilde $10^{-7}$. The Andor camera can collect entire $2048 \times 512$ pixel images at a rate of 1 Hz or full-vertical-binning (FVB) images at up to 120 Hz, necessitating integration over multiple events from the kHz rate LCLS-II beam as discussed below. A new high-throughput spherical VLS grating spectrometer is currently under commissioning as of mid-2025 and will be described in a forthcoming publication.

\begin{table}[bt]

\begin{ruledtabular}
\caption{VLS Spectrometer Parameters. Benchmarked at 540 eV, first diffraction order, a 20 $\mu$m source and a 24 $\mu$m CCD resolution.}
\label{vls-table}

\begin{tabular}{ll}
 \textbf{Overall Performance}\\
 \hline
 Resolving Power (E/$\Delta$E) & 1500\\
 Throughput, FVB & $4 \times 10^{-8}$ \\
 Throughput, Full Image & $1.6 \times10^{-7}$\\
 
 \hline
 \textbf{Spherical Mirror} & \textbf{}\\
 \hline
 Radius of Curvature (m) & 28.2\\
 Incidence Angle (°) & 2.12\\
 \hline
 \textbf{VLS Grating} & \textbf{}\\
 \hline
 Incidence Angle (°) & 1.485\\
 $D_0$ (l/mm) & 1200\\
 $D_1$ (l/mm$^2$) & 2.19\\
 $D_2$ (l/mm$^3$) & 0\\
 Blaze (°) & 1.7\\
 Coating & Au\\
 \hline
 \textbf{Detector} & Andor Newton SO\\
 \hline
 Max Readout Rate, FVB (Hz) & 120\\
 Max Readout Rate, Full (Hz) &  1 \\
 Readout Time, FVB (ms) & 5 \\
 Readout Time, Full (ms) & 392 \\
 Full Image Size (px) & $2048\times 512$ \\
 Pixel Size ($\mu$m) & $13.5\times 13.5$ \\
\end{tabular}
\end{ruledtabular}
\end{table}

\subsubsection{Avalanche Photodiodes}

\begin{figure}[bt]
    \centering
    \includegraphics[width=0.8\linewidth]{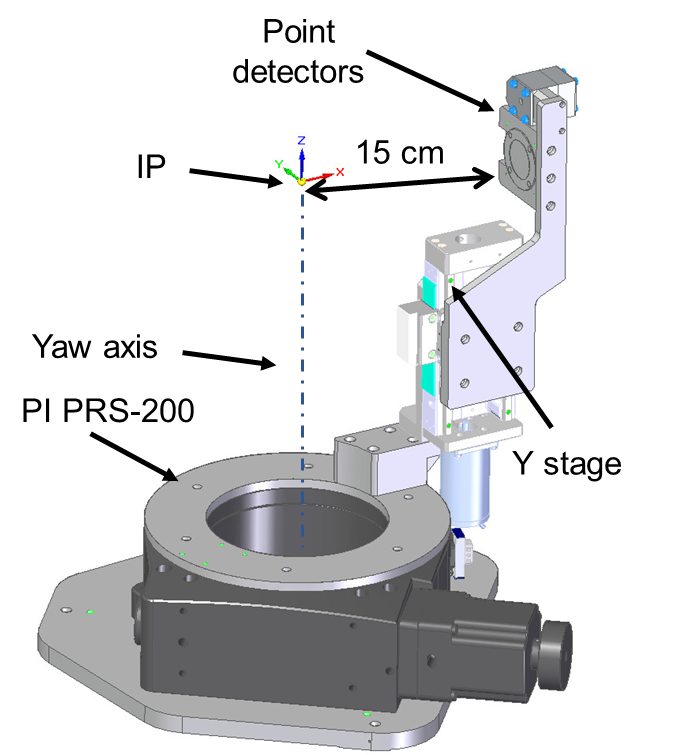}
    \caption{Model of the APD stage.}
    \label{fig:APDs}
\end{figure}

Up to three APDs (Laser Components SAR3000E1) are mounted on a stage (Physik Instruments PRS-200) which can rotate around the liquid target, ranging from a 60° to a 160° scattering angle at a distance of 15 cm from the jet. A model of the APD assembly can be seen in Fig.~\ref{fig:APDs} and associated parameters in Table~\ref{apd-table}. The APDs are mounted with thin aluminum filters that reject optical laser light while allowing most of the X-ray fluorescence photons to pass through. The foils are supported with nickel meshes to avoid damage during venting or roughing the chamber, or from debris from the liquid jet. The APDs are biased to approximately $-170$ V and the output is fed into adjustable transimpedance amplifiers (Femto DHCPA-100).

Using SLAC built Wave8-v2 digitizers (identical to those used for the FIMs\cite{heimann2025jpcs}), the APDs can be read out at up to the eventual full LCLS-II repetition rate of 929 kHz. The APDs then provide the only shot-by-shot readout in the standard liquid configuration, which also enables the best potential time resolution in pump-probe measurements as it allows for arrival time jitter between the X-rays and optical laser to be corrected on a shot-by-shot basis.

\begin{table}[bt]

\begin{ruledtabular}
\caption{APD Parameters}
\label{apd-table}

\begin{tabular}{ll}
 \textbf{APD} & \textbf{}\\
 \hline
 Part Number & Laser Components SAR3000E1\\
 Diode Material & Si \\
 Diameter (mm) & 3 \\
 Readout Rate & Up to 929 kHz \\
 Bias (V) & 160-180 \\
 \hline
 \textbf{Filter} & \textbf{}\\
 \hline
 Manufacturer & Luxel\\
 Foil & 100 nm Al | 100 nm Polyimide | 100 nm Al\\
 Mesh & 1000 lines/in square Ni grid, 7 $\mu$m wire\\
 
 \hline
 \textbf{Stage} & \textbf{}\\
 \hline
 APD Quantity & 3\\
 Distance to IP (cm) & 15\\
 Angle Range (°) & 60 to 160\\
 
\end{tabular}
\end{ruledtabular}
\end{table}

\subsubsection{Transmission XAS Detector}

A transmission XAS detector (Andor Newton SO CCD) is currently located downstream of the chemRIXS chamber for measuring the transmitted intensity of X-rays through the liquid target. A light-tight 200 nm Al filter is placed just upstream of the camera to reject optical laser light, and a pair of filter paddles with exchangeable X-ray filters (typically several $\mu$m of Co or Al, depending on photon energy) is used to attenuate the intensity of the transmitted beam to appropriate levels for the camera. As with the emission spectrometer camera, the Andor camera is limited to a 120 Hz repetition rate in FVB mode, which was sufficient for shot-by-shot detection with the LCLS-I rate, but necessitates integrating multiple events from the kHz rate beam.

\subsubsection{Transmission Spectrometer}

Preliminary transmission spectrometer measurements have been performed by moving the portable VLS spectrometer described above into the direct beam path. This configuration was used for the nonlinear X-ray experiments described in Section 3.2.2 which require the non-monochromatized "pink" beam. There are currently plans to install a high-rate transmission X-ray spectrometer based on a design developed at the TMO endstation at LCLS, which is expected to be offered for experiments in mid-2026.\cite{larsen2023oe} This spectrometer uses transmission off-axis Fresnel zone plates (FZPs) which images the beam onto a Ce:YAG scintillator that is imaged by a microscope objective and CCD detector. A similar high-rate detection scheme is also being commissioned for transmission XAS measurements.

\subsubsection{Bursting Operation}
\begin{figure}[bt]
    \centering
    \includegraphics[width=1\linewidth]{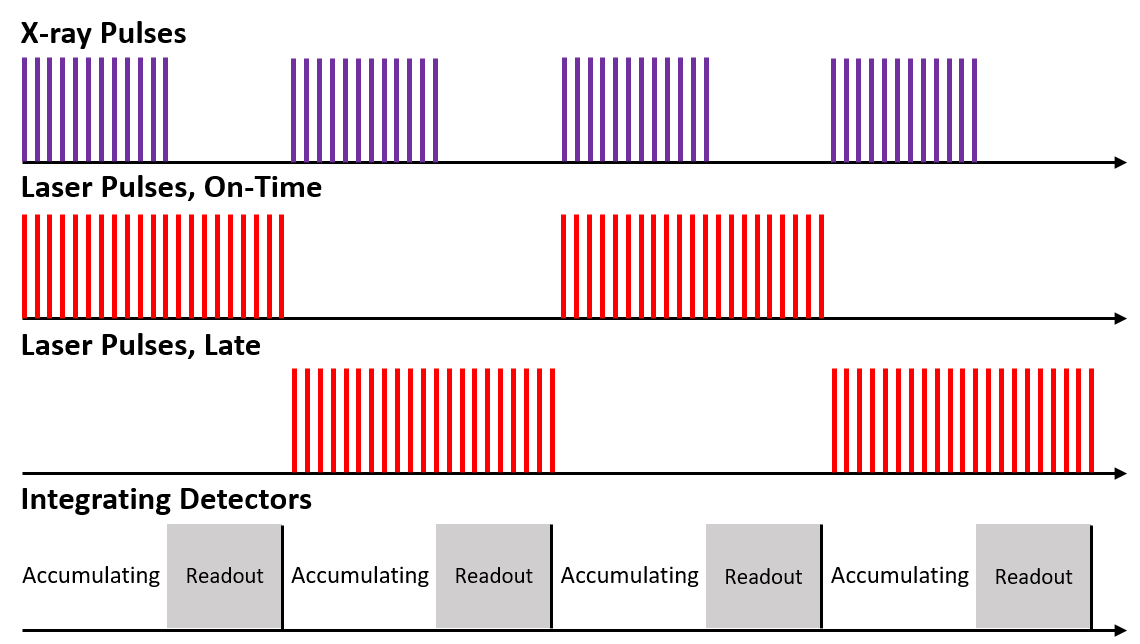} 
    \caption{Schematic of bursting operation from LCLS-II in combination with slow integrating detectors. The integrating detectors are sensitive to X-rays during their readout period, which can lead to artifacts in the data. No X-ray pulses are delivered during the detector readout period. The laser is delayed one oscillator bucket every other readout period to provide the laser-pumped and unpumped data for pump-probe measurements.}
    \label{fig:bursting}
\end{figure}

2D detectors at chemRIXS cannot operate at the full repetition rates provided by LCLS-II. If the detectors are exposed to X-rays during their readout period, artifacts can appear in the detector images and individual X-ray shots can be split between different images, complicating normalization and event filtering procedures. To avoid these issues, the XFEL is usually operated in a “bursting” mode, where the accelerator does not produce X-ray pulses during the detector readout period (which typically drops the effective rep rate by approximately \textasciitilde 33-50\%). The optical pump laser must also be synchronized to these detectors, such that each detector frame is either fully pumped or unpumped. A schematic of the typical burst pattern from the perspective of the XFEL, laser, and detector is shown in Fig.~\ref{fig:bursting}. Typical burst pattern parameters used at chemRIXS are shown in Table~\ref{burst-table}. In choosing a burst pattern, a balance has to be struck between duty cycle and the time resolution, as longer integration times result in greater arrival-time jitter which cannot be corrected using the arrival time monitor (discussed below).

\begin{table}[bt]
\begin{ruledtabular}
\caption{Typical chemRIXS LCLS-II bursting configurations, using Andor Newton SO CCD cameras in either full image or full vertical binning (FVB) mode as the integrating detector with typical accumulating times.}
\label{burst-table}

\begin{tabular}{ lrrr  }
 \textbf{Detector Property} & \textbf{Andor Image} & \textbf{Andor FVB} & \textbf{Andor FVB}\\
 \hline
 Readout Time (ms) & 396  & 5 & 5 \\
 Accumulating Time (ms) & 604 & 10 & 5\\
 Readout Rate (Hz) & 1 & 66 & 102 \\
 Duty Cycle                 & 60\%    & 66\% & 50\% \\
 \hline
 \textbf{Pulses/sec.} \\
 \hline
 33.2 kHz Base Rate & 20030 & 22108 & 16582  \\
 16.6 kHz Base Rate & 10015 & 11054 & -\textsuperscript{a}  \\
 8.3 kHz Base Rate & 5008 & 5527 & -\textsuperscript{a}  \\
 
\end{tabular}
\footnotetext{Detector rate not a subharmonic of beam rate.}
\end{ruledtabular}
\end{table}

\subsection{Optical Lasers}
\begin{table*}

\begin{ruledtabular}
\caption{Core optical laser (OPCPA) parameters by wavelength. Laser repetition rate is up to 33 kHz to match X-ray repetition rate. Typical laser spot sizes are 100-200 $\mu$m FWHM.}
\label{laser-table}

\begin{tabular}{ l r r r r r }
 \textbf{Parameter} & \textbf{800 nm} & \textbf{400 nm}& \textbf{266 nm}& \textbf{480-600 nm}& \textbf{600-900 nm}\\
 \hline

 Pulse Duration (fs) & 20 & 25 & 30 & $<$50 & $<$50\\
 Pulse Energy ($\mu$J) & 500 & 50 & 5-10 & \textasciitilde10 & \textasciitilde5\\
 Fluence (mJ/cm$^2$) & 1500 & 150 & 15-50 & 50 & 15\\

\end{tabular}
\end{ruledtabular}
\end{table*}

\subsubsection{Laser System}

For the initial LCLS-II experimental runs (through December 2025), chemRIXS used an OPCPA (optical parametric chirped pulse amplifier) to generate 1 mJ pulses centered at 800 nm with bandwidth supporting a transform limit of \textasciitilde 20 fs. The OPCPA is pumped by a 400 W InnoSlab amplifier (Amphos 3000 series) and seeded by white light generated by a Light Conversion Carbide CB3-80W. Both systems are seeded by an oscillator (Light Conversion, Flint) which is frequency locked to RF from the Linac. The system operates at 33 kHz (or subharmonics thereof) and is located in the nearby central laser hall. The pulses are transported in vacuum to the endstation where the pulses are compressed and any frequency conversion required can take place. At 800 nm, the pulses have durations <25 fs, pulse energies of 500 $\mu$J, and can be focused to a spot size of \textasciitilde 100 $\mu$m FWHM in diameter at the IP. Additional pump wavelengths can be produced using the second or third harmonic or using a visible OPA (480-900 nm). Typical pulse properties for each wavelength are shown in Table~\ref{laser-table}.

The optical lasers are in-coupled to the chemRIXS IP via an edge cut square mirror which allows the laser and X-ray pulses to be nearly collinear (crossing angle of \textasciitilde 0.5\textdegree). Depending on wavelength, the lasers can be focused to a spot of roughly 150 $\mu$m in diameter (FWHM), or astigmatically to 250 x 150 $\mu$m. Downstream of the IP, the lasers are outcoupled using the outcoupling holey lens shown in Fig.~\ref{fig:chemrixs} and a holey mirror.

Future chemRIXS runs (starting Fall 2026) will instead use ytterbium laser systems directly (fundamental wavelength of 1030 nm), which will be described in later publications.

\subsubsection{Laser/ X-ray Timing}
The time delay between the X-ray and laser pulses is controlled electronically, which allows for pulse-by-pulse control of the delay on the femtosecond to microsecond scale. The timing jitter between the XFEL and laser has been measured to be roughly 100 fs FWHM on short time scales (<1 s), although this can be compensated with the arrival time monitor (ATM) described below (depending on detector integration time). Similarly, there are diurnal drifts of several picoseconds in the synchronization fiber between the XFEL and optical laser system, which can be tracked and corrected over time using the ATM. To generate laser-on and laser-off shots for pump-probe measurements, the laser is delayed several ns either before or after the X-ray pulse. This scheme helps with the overall stability of the laser system and can minimize scattered-light artifacts on detectors. For use with the integrating detectors described in the previous section, burst patterns can be used such that each detector exposure only sees on-time or delayed shots (Fig.~\ref{fig:bursting}).

\subsubsection{Arrival Time Monitor (ATM)}

\begin{figure}[!h]
    \centering
    \includegraphics[width=0.8\linewidth]{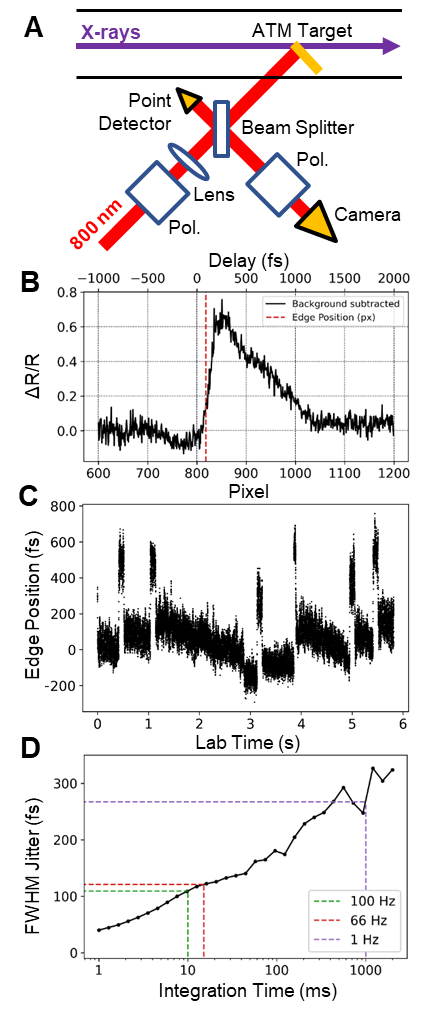}
    \caption{A. Schematic of the ATM setup. The X-ray beam is clipped on an angled target so that different parts are pumped at different times. The target is illuminated with a pickoff of the laser to observe the X-ray induced reflectivity. The onset of X-ray enhanced reflectivity is identified as time zero. B. The difference profile of the laser on the ATM target. C. Arrival time jitter as measured with the ATM. D. Plot indicating how much jitter is introduced for different integration times, limiting the time resolution of the measurement. }
    \label{fig:ATM}
\end{figure}

The arrival time monitor (ATM) is used to correct for jitter and drift between the arrival times of the X-ray and laser in pump-probe experiments.\cite{droste2020oeo,muhammad2021cle2pj} The basic ATM design can be seen in Fig.~\ref{fig:ATM}A. In its simplest form, the ATM consists of a layered semiconductor target inserted into the X-ray beam path at a 35\textdegree angle of incidence upstream of the chemRIXS IP. The target is not transmissive to the soft X-rays, so in operation it only clips the beam profile vertically (typically 50\%). The ATM target is illuminated by a pickoff of the 800 nm laser pulse at normal incidence and the reflectivity of the target is measured by a camera (Teledyne DALSA Piranha4-2k or Teledyne Adimec Opal). When the X-rays are absorbed by the semiconductor, the reflectivity of the target varies due to the presence of charge carriers in the conduction band. To maximize the optical response, we use an optimized stack of thin films with the top layer being the active semiconducting GaAs target. This stack creates an etalon effect that can greatly boost the reflectivity change by orders of magnitude, as previously discussed in the literature.\cite{droste2020oeo,muhammad2021cle2pj,eckert2015apl} Details on the optimal targets and their performance will be reported in a future publication.

As the ATM target is tilted relative to the X-ray profile, the X-ray pulse arrives to different locations of the target at different times, which allows the X-ray arrival time to be encoded spatially by the enhanced reflectivity. The ATM signal then appears as a sudden jump in reflectivity between where the laser pulse arrived before the X-ray pulse and where it arrived after the laser pulse (Fig.~\ref{fig:ATM}B). The resolution of the ATM is given by the 35\textdegree{}\ angle of incidence, which in practice gives \textasciitilde 2 fs accuracy in the edge location. The total time window of the ATM is limited by the width of the X-ray spot on the target (which is constrained by focusing requirements at the experimental IP), but is typically about 1 ps. To maintain high quality time resolution across the entire time range (and to account for changes in the laser pathlength to the chemRIXS IP between experiments), the ATM is also equipped with a mechanical delay stage which compensates for changes in the laser X-ray timing at the IP such that the timing signal stays centered on the ATM, even for delay scans beyond the ATM temporal window.

The ATM then provides information on how much the laser X-ray arrival time differs from the nominal time used during a delay scan on a shot-by-shot basis. This information can be used to correct the nominal delays for the APDs (which also report on a shot-by-shot basis) or can be used to filter and correct time delays in the integrated data. Using the ATM correction should give time resolution approaching the instrument response limit from the laser and monochromatized X-ray pulses (approx. 50-70 fs combined, depending on photon energy).

Statistics on typical raw and mean-centered arrival-time jitter are shown in Fig.~\ref{fig:ATM}C-D. From the ATM measurements, we find that the uncorrected arrival-time jitter is of order 100 fs and increases with increasing integration time, as is required for the slower detectors. For instance, a 10 ms integration time adds about 100 fs of jitter, while a 1 s integration time adds nearly 300 fs. As the ATM cannot be used to correct for integrated jitter, this provides a lower limit on the time resolution for these detection channels.

The ATM has recently been upgraded to work in an “interferometric” mode, where the unpumped reflections destructively interfere to provide an almost background-free ATM measurement.\cite{droste2020oeo} This enhancement is required for maintaining a high quality timing signal with the reduced pulse energy from monochromatized X-rays.

In practice, high quality ATM signals have been found to require at least 500 nJ of monochromatized X-rays on the target with the 300 $\mu$m X-ray spot available with typical focusing conditions at the IP. This is readily achievable at lower photon energies, but is more challenging as the accelerator performance and beamline transmission drop at higher photon energies, which makes shot-to-shot correction difficult above about 800 eV.

\subsection{Liquid Sample Delivery}
\begin{figure*}
    \centering
    \includegraphics[width=0.7\linewidth]{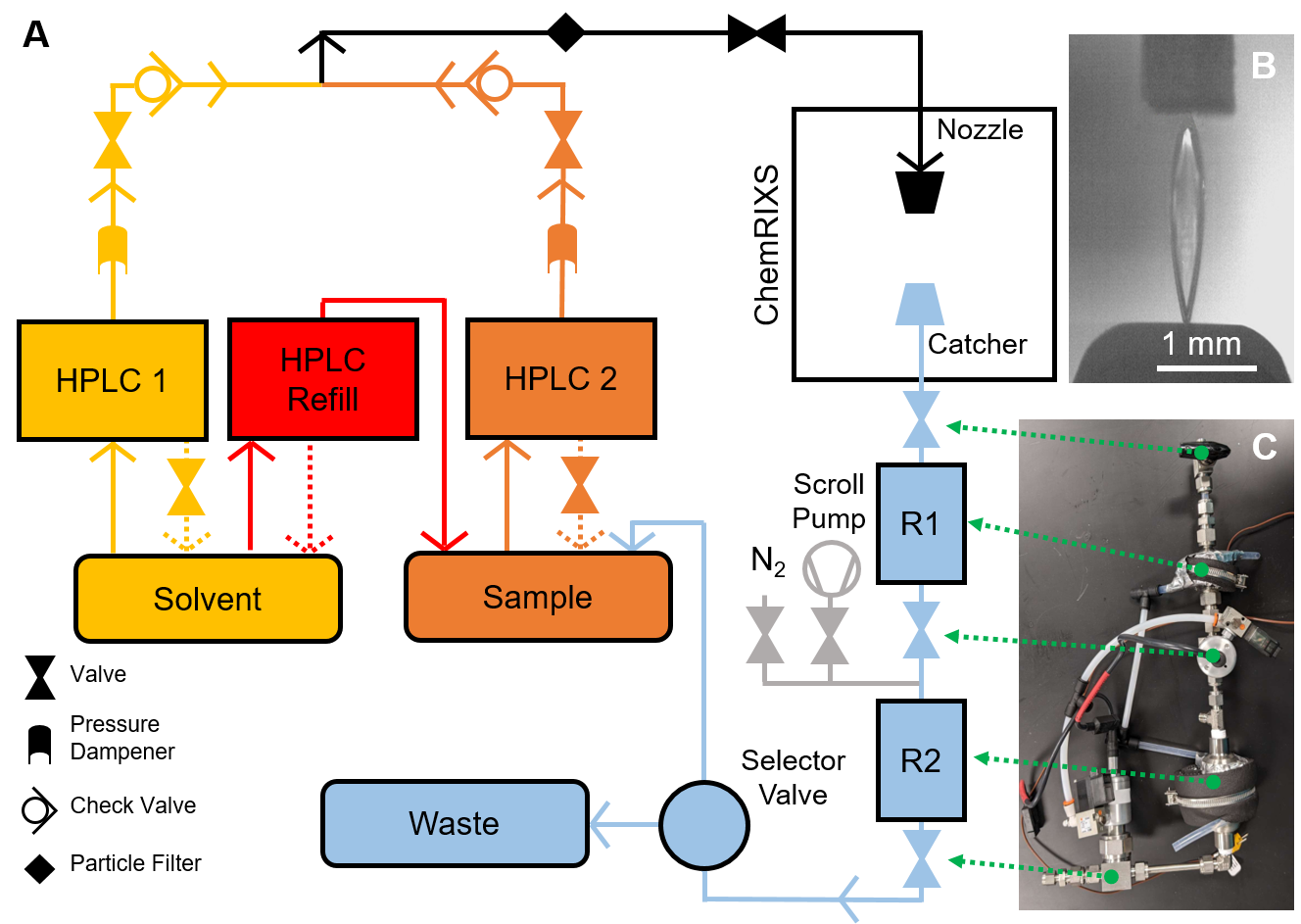}
    \caption{A. Schematic of the liquid recirculation and sample delivery system at chemRIXS. R1: recirculator collection chamber. R2: recirculator airlock chamber. B. Liquid nozzle, sheet, and catcher as seen with the inline telescope at chemRIXS. C. Picture of the core recirculator components. Dashed green arrows indicate the corresponding component in the schematic diagram.}
    \label{fig:SED}
\end{figure*}
\subsubsection{Liquid Jet Nozzles}

The core focus of the chemRIXS instrument design is soft X-ray spectroscopy on liquid phase samples. As soft X-ray spectroscopy necessitates high vacuum conditions and a focused XFEL beam would quickly destroy the windows in a liquid flow cell, the liquid samples are injected into the vacuum using liquid jets.

The main type of liquid jet used at chemRIXS is a liquid sheet produced with etched glass microfluidic nozzles.\cite{hoffman2022fmb,crissman2022lc} The liquid sheets are free-standing, large-area liquid films with thicknesses ranging from 0.2 to 10 $\mu$m depending on the nozzle dimensions. These sheets are thin enough for XAS measurements in transmission and avoid analogous saturation effects in the fluorescence measurements. The sheets are typically several mm long and several hundred $\mu$m wide (Fig.~\ref{fig:SED}B), which enables the XFEL and optical laser spots to be partially defocused if necessary to avoid damage to the sample or liquid jet structure without wasting photons by having them miss the liquid target. The liquid velocity in the sheet is >10 m/s, which ensures that sample is refreshed between each XFEL shot. The sheets also flow supercritically, which means that shockwaves in the liquid induced by the XFEL or laser shots will propagate slower than the sheet velocity. Although each XFEL shot sees a fresh liquid target, it is often still desirable to reduce average power (either focal intensity or repetition rate) to prevent build up of solid debris around the nozzle or catcher.

The sheet jet nozzles most frequently used at chemRIXS have a single tapered converging channel which makes them resistant to clogging compared to other types of capillary-based nozzles. However, these nozzles necessitate a high liquid flow rate (1.5-6 mL/min, depending on required thickness and sheet width) to both hydrodynamically form the jet and ensure the sample is replenished for each X-ray shot. The sample volumes required for the high liquid flow rates are mostly mitigated by the liquid recirculator system described below. Several other nozzle types are available as well if recirculation is not suitable for a given sample: gas accelerated liquid sheet nozzles\cite{koralek2018nc} which produce a sheet by impinging a round jet with helium (thickness 0.03-2 $\mu$m, flow rate 150-500 $\mu$L/min) or 20 $\mu$m Rayleigh jets (flow rate 150-500 $\mu$L/min).

\subsubsection{Liquid Recirculator}

To minimize the amount of sample required to run the sheet nozzles, a sample recirculation system is used to take liquid sample collected by the in-vacuum catcher back to the sample reservoir out of vacuum.\cite{hoffman2022fmb} A schematic diagram of the system is shown in Fig.~\ref{fig:SED}A.

Liquid solutions are supplied to the nozzle with commercial high pressure liquid chromatography (HPLC) pumps. The pumps are fit with pressure dampeners (a T-pipe fitting with several mL of closed volume) to remove pressure oscillations. Two HPLC pumps are always connected to the nozzle: one which is dedicated to neat solvent and one for the relevant sample. This allows for the jet to be started with neat solvent, and for sample and solvent to be rapidly toggled if necessary to clean accumulated sample debris in the IP. When recirculating small volumes, an additional HPLC pump is also used to replace solvent lost to evaporation in vacuum (of order 0.1 mL/min for water, depending on the exposed surface area of the liquid target).\cite{hoffman2022fmb}

The liquid jet enters a conical catcher with a 400-500 $\mu$m diameter hole placed a few mm below the liquid jet nozzle and the interaction point (design adapted from iRS's Jet Catcher Systems). The catcher is heated to 70-110 \textdegree C to prevent liquids from freezing to the exterior. The catcher is connected to a ¼” diameter pipe that is connected to a collection chamber ( R1 in Fig.~\ref{fig:SED}A),  which is kept under vacuum and cooled to minimize the vapor pressure of the solvent (10 \textdegree C is sufficient for water). The combination of gravity and the vapor pressure gradient between the catcher and R1 enables efficient draining from the vacuum chamber.

To cycle liquid out of the collection chamber, a valve under it is opened to drain to an airlock chamber (R2 in Fig.~\ref{fig:SED}A) which is also at rough vacuum. The valve to the collection chamber closes, and the airlock chamber is backfilled with nitrogen to pressurize it slightly above 1 atm and drain it to the sample reservoir kept at atmosphere. The airlock is then put under vacuum to prepare for the next cycle. A full cycle occurs every 60 seconds. If recirculation is not required, the output of the recirculator can be directed to a waste container instead. An image of the core recirculator components can be seen in Fig.~\ref{fig:SED}C.

Using the recirculator, sample volumes of \textasciitilde 25 mL can be used, which would normally be completely consumed within 20 minutes without recirculation with our typical flow rates. The collection chamber pressure has to be kept to $< 20$ Torr to prevent liquids from backflowing out of the catcher. The recirculator then requires solvents with relatively high boiling points to work well without modification. Thus far, water, isopropanol, dimethyl sulfoxide, dimethyl formamide, and toluene have been successfully used at chemRIXS.

\subsubsection{Vacuum System}
As liquid samples exposed to vacuum will rapidly evaporate, substantial pumping is required to maintain reasonable vacuum levels. Depending on the solvent and jet geometry/size, typical operating vacuum pressures are 0.1 - 1 mTorr during experiments. With substantial differential pumping before and after chemRIXS, this is sufficient to keep the rest of the beamline at pressures below $10^{-6}$ Torr. Most pumping is provided by four turbomolecular pumps with combined nominal pumping rates of 3800 L/s, which are sufficient to reach these target pressures when running water-based jets (and provide a base chamber pressure of $10^{-7}$ Torr). In addition, there are a pair of liquid nitrogen-fed cold traps that can freeze solvent vapor in the chamber. The cold traps can further reduce the vacuum pressure during jet operation approximately another order of magnitude. They can also be retracted and isolated from the main chemRIXS chamber, which allows them to regenerate under rough vacuum. This allows one trap to be active while the other is regenerating, and allows them to be in principle continuously cycled.

\subsection{Controls System}

The controls system comprises three integrated components: the overall beamline instrumentation, the chemRIXS endstation, and the liquid jet sample delivery system.\cite{joshi2025,zamudioestrada2025} Built on the Experimental Physics and Industrial Control System (EPICS) framework with modular programmable logic controller (PLC) based subsystems, it governs critical hardware including vacuum, motion, optics, and fluid handling, while providing standardized control interfaces, safety interlocks, and diagnostic monitoring.\cite{joshi2025} The controls support precise positioning of optical elements and sample alignment, automated experiment sequences, and real-time feedback to ensure optimal interaction with the X-ray beam. The liquid jet delivery system is fully integrated into this framework, enabling precise regulation of flow rate and pressure, continuous monitoring of jet stability, and interlocks to prevent sample loss or system faults.\cite{zamudioestrada2025} Together, these interconnected control components facilitate flexible experimental configurations and ensure reliable, high-repetition-rate performance for time-resolved experiments.

\section{Initial Scientific Results}

\begin{figure}[bt]
    \centering
    \includegraphics[width=0.9\linewidth]{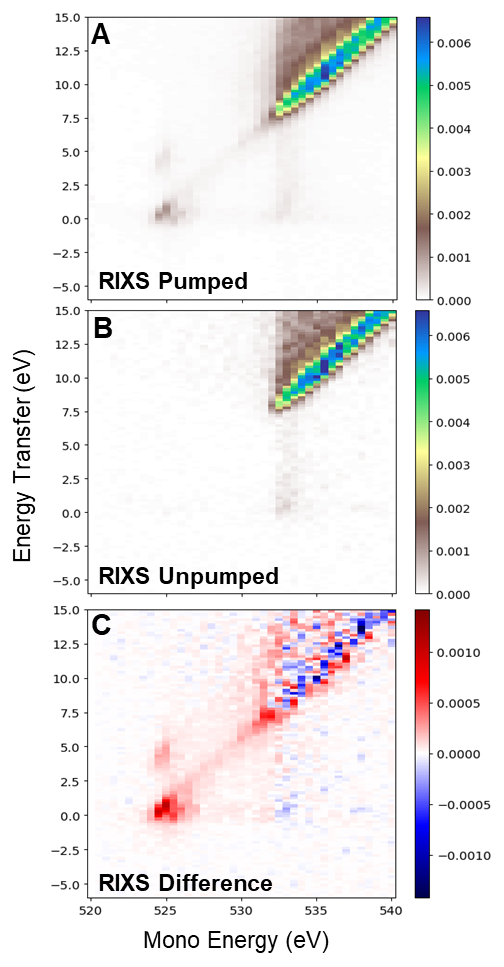}
    \caption{RIXS maps of water before (A) and 2 ps after strong field ionization (B). A difference RIXS map is shown in C. The scan was collected in 5 minutes at a 5.5 kHz repetition rate. The water valence hole (525 eV) and solvated electron feature (532 eV) are clearly visible in the RIXS difference map.}
    \label{fig:WaterRIXS}
\end{figure}


\subsection{LCLS-II Commissioning Results}

At the conclusion of the LCLS-II chemRIXS commissioning program in June 2024, a validation experiment was performed on the strong-field ionization of water. This experiment is analogous to those performed previously at LCLS-I at both the SXR and chemRIXS instruments.\cite{kjellsson2020prl,loh2020s,sopenamoros2024jacs} Strong-field ionization results in an excess electron and an ionized water species which undergoes proton transfer with an adjacent water molecule to yield an OH radical and a hydronium (H$_3$O$^+$) ion. This results in two transient absorption signatures in the oxygen K-edge XAS spectra. The RIXS maps also show the couplings between the newly created valence hole and lower-lying occupied electronic orbitals.

In this experiment, a strong 800 nm pump laser (200 $\mu$J with a 120 $\mu$m FWHM spot size) impinged on a 2 $\mu$m thick water sheet jet. The X-ray pulses were delivered at an 8.3 kHz repetition rate (using a 66 Hz burst pattern that reduced the effective rate to about 5.5 kHz, see Table~\ref{burst-table}) with a pulse energy of 250 $\mu$J, a 200 $\times$ 70 $\mu$m FWHM spot size, and monochromatized to a bandwidth of 300 meV, corresponding to about 2 $\mu$J of pulse energy delivered to the IP.

\begin{figure}[bt]
    \centering
    \includegraphics[width=0.9\linewidth]{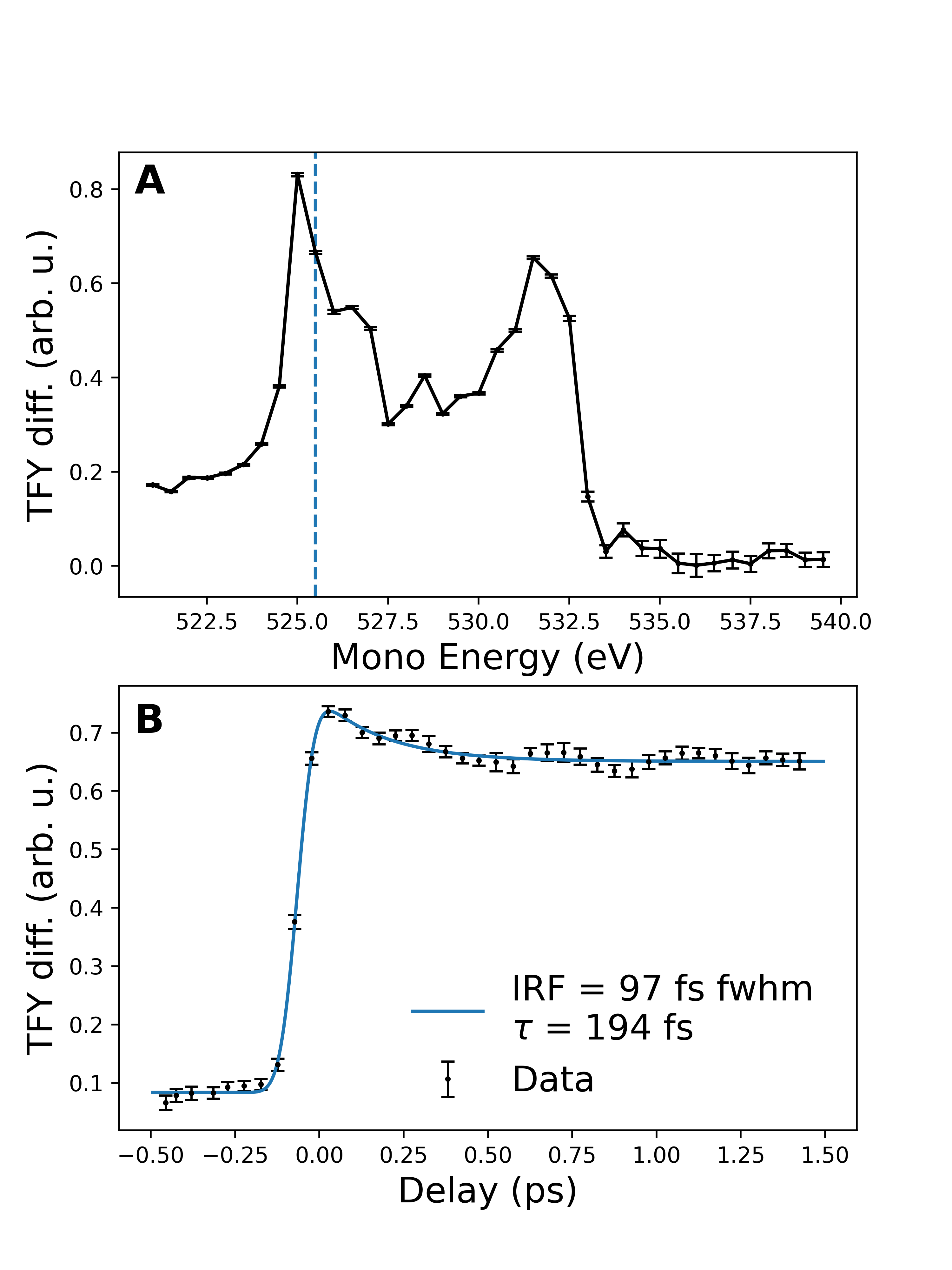}
    \caption{A) Difference signal from strong field ionized water TFY-XAS at a 2 ps time delay, measured concurrently with the RIXS maps in Fig.~\ref{fig:WaterRIXS}. B) Time delay scan of the valence hole feature at 525.5 eV in strong field ionized water measured with TFY-XAS after ATM correction. The error bars are estimated from the statistical uncertainty in each time bin and correspond to $\pm 2 \sigma$. Best fit of the instrument response gives a <100 fs resolution FWHM. The $\sim 200$ fs decay feature that has been previously reported\cite{loh2020s} is also apparent.}
    \label{fig:WaterDelay}
\end{figure}

The RIXS results are shown in Fig.~\ref{fig:WaterRIXS}. These measurements are a result of a single 5 minute scan using the high LCLS-II repetition rate. Pumped and unpumped RIXS maps are shown in panels A and B at a 2 ps delay, with the difference RIXS map shown in panel C. The simultaneously-measured difference TFY-XAS can be seen in Fig.~\ref{fig:WaterDelay} A. The unpumped RIXS map demonstrates the known oxygen edge resonances of liquid water, which is primarily the $1\mathrm{b}_1 \rightarrow 1\mathrm{a}_1$ emission peak for input photon energies above 533 eV.\cite{nilsson2013jesrp} As this feature emits at near constant energy with respect to input energy, it presents as a diagonal feature in the RIXS map (where energy transfer is input photon energy minus emitted photon energy).

The signature of the OH radical species is immediately obvious in all difference measurements, appearing as the strong transient absorption around 525 eV due to the ejection of electrons from the $1\mathrm{b}_1$ water valence orbital.\cite{loh2020s} A core-excited electron can refill this "valence hole", which creates a low-energy transient absorption. In the RIXS data, in addition to the elastic scattering at zero energy transfer, the previously reported 4 eV energy transfer feature is also apparent, corresponding to the energy difference between the excited $1\mathrm{s} \rightarrow 1\pi$ transition and the emitted $2\sigma \rightarrow 1s$ transition in the OH radical.\cite{kjellsson2020prl} Compared to the previously published time-resolved RIXS maps on this system, we were able to acquire a dataset with 3 times the statistics (>1 million events) in 1/20th the time.\cite{kjellsson2020prl}

An example time-delay TFY-XAS scan on the water valence hole feature (525.5 eV) is shown in Fig.~\ref{fig:WaterDelay} B after using the ATM to correct for the timing drifts, jumps, and jitter shown in Fig.~\ref{fig:ATM}. Best fit of the data gives an instrument response function of 97 $\pm 25$ fs FWHM, assuming an impulsive response. A fast $\sim$ 200 fs decay process is also visible, which has been previously assigned to the vibrational cooling of the OH radical product.\cite{loh2020s}

\subsection{LCLS-I Results}

\subsubsection{Time-Resolved XAS}

While the low repetition rate from LCLS-I is not ideal for time-resolved RIXS measurements, it is still sufficient for time-resolved XAS measurements. In addition to the prior measurement examining the strong-field ionization of water,\cite{sopenamoros2024jacs} chemRIXS was used to study metal-ligand charge transfer in derivatives of the ruthenium-based photosensitizer Tris(bipyridine)ruthenium(II) chloride (Rubipy) using nitrogen K-edge XAS.\cite{ryland2025ac,kahraman2025} By replacing one of the bipyridine ligands with a larger dipyridophenazine (dppz) ligand, the charge transfer was found to be localized on the extremal components of the new ligand. Two mechanisms were found for this transfer: a newly-indentified sub-70 fs intra-ligand transfer within the dppz moiety and a 2 ps hopping mechanism from the neighboring bipyridine ligands.\cite{ryland2025ac} By instead replacing one of the bipyridine ligands with a proton-accepting bipyrazine ligand, the initial steps of proton-coupled electron transfer were observed in acidic solutions with site-specificity.\cite{kahraman2025}

\subsubsection{Nonlinear Soft X-ray Spectroscopies}

The development of broadband attosecond X-ray pulses\cite{duris2020np,franz2024np} and two-color pulse pairs \cite{guo2024np} from LCLS have also enabled a number of nonlinear X-ray spectroscopy methods at chemRIXS. As these experiments require taking the full, unmonochromatized beam, they have thus far been measured using the portable VLS spectrometer in the transmission geometry (using sufficiently thin liquid sheets). As was discussed above, the permanent transmission FZP spectrometer\cite{larsen2023oe} is also presently being commissioned. Thus far, these experiments have only been performed using the LCLS-I beam, although they will be offered with the LCLS-II beam in the future.

Two-color all X-ray attosecond transient absorption spectroscopy (AX-ATAS) was performed to investigate the radiolysis of water on subfemtosecond timescales, where an X-ray pump pulse at 260 eV was used to generate valence holes that were probed by an X-ray probe pulse at 525 eV with a 0.6 fs delay.\cite{li2024s} The results showed that the splitting observed in the water X-ray emission spectra (XES) arises from ultrafast motions in ionized water and not different structural motifs, as here the transient XAS gives the inverse of the XES signal. The broad bandwidth and coherence of the attosecond pulse was used for impulsive stimulated X-ray Raman scattering (ISXRS) at the oxygen K-edge in water, where a transition is driven with the blue edge of the pulse spectrum and a stokes emission feature is stimulated at the red edge of the spectrum.\cite{alexander2024sa} ISXRS creates coherences between valence states, allowing insights into electronic couplings. Soft X-ray second harmonic generation (SXSHG) was used to measure a surface-selective spectrum of water by taking advantage of the high peak intensities of the attosecond X-ray pulses and selection rules which eliminate the signal from bulk water.\cite{hoffman2025nca} The SXSHG spectrum showed a shift relative to the bulk XAS which was associated with broken hydrogen bond structures at the water surface.

\subsubsection{Solid State Measurements}

While the dedicated qRIXS materials science instrument was still under construction, chemRIXS was modified to enable solid state measurements. The experimental setup has been described in a previous publication.\cite{assefa2022rsi} Briefly, the sample was mounted using the cryostat manipulator mentioned above. An electromagnet was mounted to another manipulator and a camera with a controllable mask was mounted in a forward-scattering geometry.

X-ray photon correlation spectroscopy (XPCS) and its ultrafast analogue, X-ray photon fluctuation spectroscopy (XPFS),\cite{shen2021ma} were performed to examine magnetic ordering in quantum materials.\cite{plumley2024apx} These techniques take advantage of the high degree of coherence of XFEL pulses to produce scattered speckle patterns. The correlations of these speckle patterns are measured over time to monitor the dynamics in the material. Using XPFS, the phase boundary between stripe and skyrmion magnetic domains was identified in amorphous FeGd films.\cite{plumley2024apx} In an amorphous FeGe system, the distinct timescales for dynamics in two different nematic magnetic phases were identified with XPFS, as well as a change in the reorientation of the nematic state.\cite{tumbleson2025sa}

\section{Summary}

The chemRIXS instrument offers many new opportunities for studying solution-phase systems with time-resolved soft X-ray spectroscopy. The high rep-rate from LCLS-II combined with advances in the detectors, optical laser system, and liquid jet recirculation system will enable studies on dilute systems. These developments open up time-resolved XAS and RIXS to entirely new classes of samples, as well as enabling the development of new soft X-ray spectroscopies on liquid targets. More details on the chemRIXS instrument with current operating parameters can be found at https://lcls.slac.stanford.edu/instruments/chemrixs.

\section{Facility Access}

LCLS instruments are open to academia, industry, government agencies, and research institutes worldwide for scientific investigations. There are two calls for proposals per year and an external peer-review committee evaluates proposals based on scientific merit and instrument suitability. Access is without charge for users who intend to publish their results. Prospective users are encouraged to contact instrument staff members to learn more about the science and capabilities of the facility, and opportunities for collaboration.

\begin{acknowledgments}
This research was carried out at the Linac Coherent Light Source (LCLS) at the SLAC National Accelerator Laboratory. SLAC National Accelerator Laboratory is supported by the US Department of Energy, Office of Science, Office of Basic Energy Sciences under Contract No. DE-AC0276SF00515.

 B.I.P. was supported by the U.S. Department of Energy, Office of Science, Office of Basic Energy Sciences, Chemical Sciences, Geosciences and Biosciences Division. S.L., E.P., and L.Y. were supported by the US Department of Energy, Office of Science, Basic Energy Science, Chemical Sciences, Geosciences and Biosciences Division under Contract No. DE-AC02-06CH11357. A.N. and M.K. are supported by the US Department of Energy, Office of Science, Basic Energy Science under Contract No. DE-SC0023249. A.N. is additionally supported by the Wenner-Gren foundation.

We acknowledge the entire LCLS staff, as well as the SLAC vacuum, metrology, and machine shops staff, for their assistance during the design, construction, and commissioning of the chemRIXS instrument, including M. H. Seaberg, H. Wang, Q. Yougoubare, P. Oppermann, S. Guillet, J. Zamudio Estrada, A. Liang, C. Pino, and R. Lodico.
\end{acknowledgments}

\section*{Author Declarations}

\subsection*{Conflict of Interest}

The authors have no conflicts to disclose.

\subsection*{Author Contributions}

\textbf{David J. Hoffman}: Investigation (equal), Visualization (lead), Writing/Original Draft Preparation (lead), Writing/Review \& Editing (lead). \textbf{Douglas Garratt}: Investigation (equal), Formal Analysis (equal), Visualization (equal), Writing/Original Draft Preparation (equal). \textbf{Matthew Bain}: Investigation (equal), Writing/Original Draft Preparation (equal). \textbf{Christina Y. Hampton}: Investigation (equal), Visualization (equal), Resources (equal). \textbf{Benjamin I. Poulter}: Investigation (equal), Formal Analysis (equal), Visualization (equal), Writing/Original Draft Preparation (equal). \textbf{Jyoti Joshi}:  Resources (lead), Writing/Original Draft Preparation (equal). \textbf{Giacomo Coslovich}: Investigation (equal), Writing/Review \& Editing (equal). \textbf{Frank P. O’Dowd}: Resources (lead), Project Administration (equal). \textbf{Daniel P. DePonte}: Investigation (equal), Resources (equal), Writing/Review \& Editing (equal). \textbf{Alexander H. Reid}: Investigation (equal). \textbf{Lingjia Shen}: Investigation (equal). \textbf{Daniel Jost}: Investigation (equal). \textbf{Mina R. Bionta}: Investigation (equal), Writing/Review \& Editing (equal). \textbf{Joshua J. Turner}: Investigation (equal), Writing/Review \& Editing (equal). \textbf{Ming-Fu Lin}: Investigation (equal). \textbf{Philip Heimann}: Investigation (equal), Resources (equal). \textbf{Stefan P. Moeller}: Investigation (equal), Resources (equal). \textbf{Jake D. Koralek}: Investigation (equal).
\textbf{Tyler Johnson}: Resources (equal). \textbf{K Ninh}: Resources (equal), Project Administration (equal). \textbf{Raybel Almeida}: Resources (equal), Project Administration (equal), Writing/Review \& Editing (equal). \textbf{Adam Berges}: Resources (equal). \textbf{Stephanie Fung}: Resources (equal). \textbf{Shuai Li}: Investigation (equal), Formal Analysis (equal), Writing/Review \& Editing (equal). \textbf{Eetu Pelimanni}: Investigation (equal), Formal Analysis (equal), Writing/Review \& Editing (equal). \textbf{Amke Nimmrich}: Investigation (equal), Formal Analysis (equal), Writing/Review \& Editing (equal). \textbf{Munira Khalil}: Supervision (equal), Writing/Review \& Editing (equal). \textbf{Linda Young}: Supervision (equal), Writing/Review \& Editing (equal). \textbf{Thomas J.A. Wolf}: Conceptualization (equal), Supervision (equal), Investigation (equal), Project Administration (equal), Writing/Review \& Editing (equal). \textbf{Kristjan Kunnus}: Conceptualization (equal), Investigation (lead), Project Administration (equal), Visualization (equal), Writing/Original Draft Preparation (equal). \textbf{Georgi L. Dakovski}: Conceptualization (lead), Investigation (equal), Project Administration (lead), Supervision (equal), Writing/Original Draft Preparation (equal).

\section*{Data Availability Statement}
The data that support the findings of
this study are available from the
corresponding author upon reasonable
request.

\bibliography{chemrixs_instrument}

@article{alexander2024sa,
  title = {Attosecond Impulsive Stimulated {{X-ray Raman}} Scattering in Liquid Water},
  author = {Alexander, Oliver and Egun, Felix and Rego, Laura and Gutierrez, Ana Martinez and Garratt, Douglas and C{\'a}rdenas, Gustavo Adolfo and Nogueira, Juan J. and Lee, Jacob P. and Zhao, Kaixiang and Wang, Ru-Pan and Ayuso, David and Barnard, Jonathan C. T. and Beauvarlet, Sandra and Bucksbaum, Philip H. and Cesar, David and Coffee, Ryan and Duris, Joseph and Frasinski, Leszek J. and Huse, Nils and Kowalczyk, Katarzyna M. and Larsen, Kirk A. and Matthews, Mary and Mukamel, Shaul and O'Neal, Jordan T. and Penfold, Thomas and Thierstein, Emily and Tisch, John W. G. and Turner, James R. and Vogwell, Josh and Driver, Taran and Berrah, Nora and Lin, Ming-Fu and Dakovski, Georgi L. and Moeller, Stefan P. and Cryan, James P. and Marinelli, Agostino and Pic{\'o}n, Antonio and Marangos, Jonathan P.},
  year = {2024},
  month = sep,
  journal = {Science Advances},
  volume = {10},
  number = {39},
  pages = {eadp0841},
  issn = {2375-2548},
  doi = {10.1126/sciadv.adp0841},
  pmid = {39321305}
}

@article{chuang2017rosi,
  title = {Modular Soft X-Ray Spectrometer for Applications in Energy Sciences and Quantum Materials},
  author = {Chuang, Yi-De and Shao, Yu-Cheng and Cruz, Alejandro and Hanzel, Kelly and Brown, Adam and Frano, Alex and Qiao, Ruimin and Smith, Brian and Domning, Edward and Huang, Shih-Wen and Wray, L. Andrew and Lee, Wei-Sheng and Shen, Zhi-Xun and Devereaux, Thomas P. and Chiou, Jaw-Wern and Pong, Way-Faung and Yashchuk, Valeriy V. and Gullikson, Eric and Reininger, Ruben and Yang, Wanli and Guo, Jinghua and Duarte, Robert and Hussain, Zahid},
  year = {2017},
  month = jan,
  journal = {Review of Scientific Instruments},
  volume = {88},
  number = {1},
  pages = {013110},
  issn = {0034-6748},
  doi = {10.1063/1.4974356},
  urldate = {2025-06-28}
}

@article{crissman2022lc,
  title = {Sub-Micron Thick Liquid Sheets Produced by Isotropically Etched Glass Nozzles},
  author = {Crissman, Christopher J. and Mo, Mianzhen and Chen, Zhijiang and Yang, Jie and Huyke, Diego A. and Glenzer, Siegfried H. and Ledbetter, Kathryn and F. Nunes, J. Pedro and Ng, May Ling and Wang, Hengzi and Shen, Xiaozhe and Wang, Xijie and DePonte, Daniel P.},
  year = {2022},
  journal = {Lab on a Chip},
  volume = {22},
  number = {7},
  pages = {1365--1373},
  publisher = {Royal Society of Chemistry},
  issn = {1473-0197},
  doi = {10.1039/D1LC00757B},
  pmid = {35234235}
}

@article{dakovski2015jsr,
  title = {The Soft {{X-ray}} Research Instrument at the Linac Coherent Light Source},
  author = {Dakovski, Georgi L. and Heimann, Philip and Holmes, Michael and Krupin, Oleg and Minitti, Michael P. and Mitra, Ankush and Moeller, Stefan and Rowen, Michael and Schlotter, William F. and Turner, Joshua J.},
  year = {2015},
  month = may,
  journal = {Journal of Synchrotron Radiation},
  volume = {22},
  pages = {498--502},
  publisher = {International Union of Crystallography},
  issn = {16005775},
  doi = {10.1107/S160057751500301X}
}

@article{droste2020oeo,
  title = {High-Sensitivity x-Ray/Optical Cross-Correlator for next Generation Free-Electron Lasers},
  author = {Droste, Stefan and Zohar, Sioan and Shen, Lingjia and White, Vaughn E. and {Diaz-Jacobo}, Elizabeth and Coffee, Ryan N. and Reid, Alexander H. and Tavella, Franz and Minitti, Michael P. and Turner, Joshua J. and Robinson, Joseph S. and Fry, Alan R. and Coslovich, Giacomo},
  year = {2020},
  month = aug,
  journal = {Optics Express},
  volume = {28},
  number = {16},
  pages = {23545--23553},
  publisher = {Optica Publishing Group},
  issn = {1094-4087},
  doi = {10.1364/OE.398048},
  urldate = {2025-06-30},
  copyright = {{\copyright} 2020 Optical Society of America},
  langid = {english}
}

@article{heimann2019jsr,
  title = {Fluorescence Intensity Monitors as Intensity and Beam-Position Diagnostics for {{X-ray}} Free-Electron Lasers},
  author = {Heimann, P. and Reid, A. and Feng, Y. and Fritz, D.},
  year = {2019},
  month = mar,
  journal = {Journal of Synchrotron Radiation},
  volume = {26},
  number = {2},
  pages = {358--362},
  publisher = {International Union of Crystallography},
  issn = {1600-5775},
  doi = {10.1107/S1600577519001802},
  urldate = {2025-06-29},
  langid = {english}
}

@article{heimann2025jpcs,
  title = {Characterization of the {{LCLS-II X-rays}} with {{Imagers}}, {{Power Meters}} and {{Fluorescence Intensity Monitors}}},
  author = {Heimann, Philip and Alexander, Oliver and Cryan, James and Dakovski, Georgi and Ding, Yuantao and Esposito, Vincent and Garratt, Douglas and Kunnus, Kristjan and Lin, Ming-Fu and Lutman, Alberto and Moeller, Stefan and Nakahara, Kazutaka and Nuhn, Heinz-Dieter and Obaid, Razib and Seaberg, Matt and Schlotter, William},
  year = {2025},
  month = may,
  journal = {Journal of Physics: Conference Series},
  volume = {3010},
  number = {1},
  pages = {012100},
  publisher = {IOP Publishing},
  issn = {1742-6596},
  doi = {10.1088/1742-6596/3010/1/012100},
  urldate = {2025-06-25},
  langid = {english}
}

@article{hettrick1986aov2i2p4,
  title = {Stigmatic High Throughput Monochromator for Soft x Rays},
  author = {Hettrick, Michael C. and Underwood, James H.},
  year = {1986},
  month = dec,
  journal = {Applied Optics, Vol. 25, Issue 23, pp. 4228-4231},
  publisher = {Optica Publishing Group},
  doi = {10.1364/AO.25.004228},
  urldate = {2025-06-29},
  copyright = {{\copyright} 1986 Optical Society of America},
  langid = {english}
}

@article{degroot2024nrmp,
  title = {Resonant Inelastic {{X-ray}} Scattering},
  author = {{de Groot}, Frank M. F. and Haverkort, Maurits W. and Elnaggar, Hebatalla and Juhin, Am{\'e}lie and Zhou, Ke-Jin and Glatzel, Pieter},
  year = {2024},
  month = jul,
  journal = {Nature Reviews Methods Primers},
  volume = {4},
  number = {1},
  pages = {45},
  publisher = {Nature Publishing Group},
  issn = {2662-8449},
  doi = {10.1038/s43586-024-00322-6},
  urldate = {2025-09-11},
  copyright = {2024 Springer Nature Limited},
  langid = {english}
}

@article{hoffman2022fmb,
  title = {Microfluidic Liquid Sheets as Large-Area Targets for High Repetition {{XFELs}}},
  author = {Hoffman, David J. and Van Driel, Tim B. and Kroll, Thomas and Crissman, Christopher J. and Ryland, Elizabeth S. and Nelson, Kacie J. and Cordones, Amy A. and Koralek, Jake D. and DePonte, Daniel P.},
  year = {2022},
  month = dec,
  journal = {Frontiers in Molecular Biosciences},
  volume = {9},
  number = {December},
  pages = {1--15},
  issn = {2296-889X},
  doi = {10.3389/fmolb.2022.1048932}
}

@article{jay2022arpc,
  title = {Capturing {{Atom-Specific Electronic Structural Dynamics}} of {{Transition-Metal Complexes}} with {{Ultrafast Soft X-Ray Spectroscopy}}},
  author = {Jay, Raphael M and Kunnus, Kristjan and Wernet, Philippe and Gaffney, Kelly J},
  year = {2022},
  month = apr,
  journal = {Annual Review of Physical Chemistry},
  volume = {73},
  number = {1},
  pages = {1--22},
  issn = {0066-426X},
  doi = {10.1146/annurev-physchem-082820-020236}
}

@article{kjellsson2020prl,
  title = {Resonant {{Inelastic X-Ray Scattering Reveals Hidden Local Transitions}} of the {{Aqueous OH Radical}}},
  author = {Kjellsson, L. and Nanda, K. D. and Rubensson, J.-E. and Doumy, G. and Southworth, S. H. and Ho, P. J. and March, A. M. and Al Haddad, A. and Kumagai, Y. and Tu, M.-F. and Schaller, R. D. and Debnath, T. and Bin Mohd Yusof, M. S. and Arnold, C. and Schlotter, W. F. and Moeller, S. and Coslovich, G. and Koralek, J. D. and Minitti, M. P. and Vidal, M. L. and Simon, M. and Santra, R. and Loh, Z.-H. and Coriani, S. and Krylov, A. I. and Young, L.},
  year = {2020},
  month = jun,
  journal = {Physical Review Letters},
  volume = {124},
  number = {23},
  pages = {236001},
  publisher = {American Physical Society},
  doi = {10.1103/PhysRevLett.124.236001},
  urldate = {2025-07-02}
}

@article{koralek2018nc,
  title = {Generation and Characterization of Ultrathin Free-Flowing Liquid Sheets},
  author = {Koralek, Jake D. and Kim, Jongjin B. and Br{\r u}{\v z}a, Petr and Curry, Chandra B. and Chen, Zhijiang and Bechtel, Hans A. and Cordones, Amy A. and Sperling, Philipp and Toleikis, Sven and Kern, Jan F. and Moeller, Stefan P. and Glenzer, Siegfried H. and DePonte, Daniel P.},
  year = {2018},
  month = dec,
  journal = {Nature Communications},
  volume = {9},
  number = {1},
  pages = {1353},
  issn = {2041-1723},
  doi = {10.1038/s41467-018-03696-w},
  pmid = {29636445}
}

@article{kunnus2012rosi,
  title = {A Setup for Resonant Inelastic Soft X-Ray Scattering on Liquids at Free Electron Laser Light Sources},
  author = {Kunnus, Kristjan and Rajkovic, Ivan and Schreck, Simon and Quevedo, Wilson and Eckert, Sebastian and Beye, Martin and Suljoti, Edlira and Weniger, Christian and Kalus, Christian and Gr{\"u}bel, Sebastian and Scholz, Mirko and Nordlund, Dennis and Zhang, Wenkai and Hartsock, Robert W. and Gaffney, Kelly J. and Schlotter, William F. and Turner, Joshua J. and Kennedy, Brian and Hennies, Franz and Techert, Simone and Wernet, Philippe and F{\"o}hlisch, Alexander},
  year = {2012},
  month = dec,
  journal = {Review of Scientific Instruments},
  volume = {83},
  number = {12},
  pages = {123109},
  issn = {0034-6748},
  doi = {10.1063/1.4772685},
  urldate = {2025-06-28}
}

@article{li2024s,
  title = {Attosecond-Pump Attosecond-Probe x-Ray Spectroscopy of Liquid Water},
  author = {Li, Shuai and Lu, Lixin and Bhattacharyya, Swarnendu and Pearce, Carolyn and Li, Kai and Nienhuis, Emily T. and Doumy, Gilles and Schaller, R. D. and Moeller, S. and Lin, M.-F. and Dakovski, G. and Hoffman, D. J. and Garratt, D. and Larsen, Kirk A. and Koralek, J. D. and Hampton, C. Y. and Cesar, D. and Duris, Joseph and Zhang, Z. and Sudar, Nicholas and Cryan, James P. and Marinelli, A. and Li, Xiaosong and Inhester, Ludger and Santra, Robin and Young, Linda},
  year = {2024},
  month = mar,
  journal = {Science},
  volume = {383},
  number = {6687},
  pages = {1118--1122},
  issn = {0036-8075},
  doi = {10.1126/science.adn6059}
}

@article{loh2020s,
  title = {Observation of the Fastest Chemical Processes in the Radiolysis of Water},
  author = {Loh, Z.-H. and Doumy, G. and Arnold, C. and Kjellsson, L. and Southworth, S. H. and Al Haddad, A. and Kumagai, Y. and Tu, M.-F. and Ho, P. J. and March, A. M. and Schaller, R. D. and Bin Mohd Yusof, M. S. and Debnath, T. and Simon, M. and Welsch, R. and Inhester, L. and Khalili, K. and Nanda, K. and Krylov, A. I. and Moeller, S. and Coslovich, G. and Koralek, J. and Minitti, M. P. and Schlotter, W. F. and Rubensson, J.-E. and Santra, R. and Young, L.},
  year = {2020},
  month = jan,
  journal = {Science},
  volume = {367},
  number = {6474},
  pages = {179--182},
  issn = {0036-8075},
  doi = {10.1126/science.aaz4740},
  pmid = {31919219}
}

@inproceedings{muhammad2021cle2pj,
  title = {Arrival {{Time Monitor}} for {{Sub-10}} Fs {{Soft X-ray}} and 800 Nm {{Optical Pulses}}},
  booktitle = {Conference on {{Lasers}} and {{Electro-Optics}} (2021), Paper {{JTu3A}}.24},
  author = {Muhammad, Isa Shams and Frimpong, Benson and Daafour, Joseph and Xu, Xiaoqing and Walter, Peter and Wolf, Thomas J. A. and Cryan, James P. and Glownia, James M. and Robinson, Joseph S. and Droste, Stefan and Coslovich, Giacomo},
  year = {2021},
  month = may,
  pages = {JTu3A.24},
  publisher = {Optica Publishing Group},
  doi = {10.1364/CLEO_AT.2021.JTu3A.24},
  urldate = {2025-06-29},
  copyright = {{\copyright} 2021 The Author(s)},
  langid = {english}
}

@article{nicolas2022p,
  title = {A {{Tunable Resolution Grating Monochromator}} and the {{Quest}} for {{Transform Limited Pulses}}},
  author = {Nicolas, Josep and Cocco, Daniele},
  year = {2022},
  month = jun,
  journal = {Photonics},
  volume = {9},
  number = {6},
  pages = {367},
  publisher = {Multidisciplinary Digital Publishing Institute},
  issn = {2304-6732},
  doi = {10.3390/photonics9060367},
  urldate = {2025-06-27},
  copyright = {http://creativecommons.org/licenses/by/3.0/},
  langid = {english}
}

@article{odowd2018pmesrei,
  title = {Engineering {{Challenges}} for the {{NEH2}}.2 {{Beamline}} at {{LCLS-II}}},
  author = {O'Dowd, Francis and Cocco, Daniele and Dakovski, Georgi and Defever, Jim and Guillet, Serge and Hardin, Corey and Morton, Daniel and Osier, Ted and Owens, Maceo and Rich, David and Zhang, Lin},
  editor = {RW (Ed.), Volker, Schaa and Keihan (Ed.), Tavakoli and Manuel (Ed.), Tilmont},
  year = {2018},
  journal = {Proceedings of the Mechanical Eng.{\textasciitilde}Design of Synchrotron Radiation Equipment and Instrumentation},
  volume = {MEDSI2018},
  pages = {3 pages, 0.957 MB},
  publisher = {JACoW Publishing, Geneva, Switzerland},
  doi = {10.18429/JACOW-MEDSI2018-THPH36},
  urldate = {2025-06-27},
  copyright = {CC 3.0},
  isbn = {9783954502073},
  langid = {english}
}

@article{plumley2024apx,
  title = {On Ultrafast X-Ray Scattering Methods for Magnetism},
  author = {Plumley, R. and Chitturi, S. R. and Peng, C. and Assefa, T. A. and Burdet, N. and Shen, L. and Chen, Z. and Reid, A. H. and Dakovski, G. L. and Seaberg, M. H. and O'Dowd, F. and Montoya, S. A. and Chen, H. and Okullo, A. and Mardanya, S. and Kevan, S. D. and Fischer, P. and Fullerton, E. E. and Sinha, S. K. and Colocho, W. and Lutman, A. and Decker, F.-J. and Roy, S. and Fujioka, J. and Tokura, Y. and Minitti, M. P. and Johnson, J. A. and Hoffmann, M. and Amoo, M. E. and Feiguin, A. and Yoon, C. and Thayer, J. and Nashed, Y. and Jia, C. and Bansil, A. and Chowdhury, S. and Lindenberg, A. M. and Dunne, M. and Blackburn, E. and Turner, J. J.},
  year = {2024},
  month = dec,
  journal = {Advances in Physics: X},
  volume = {9},
  number = {1},
  pages = {2423935},
  publisher = {Taylor \& Francis},
  issn = {null},
  doi = {10.1080/23746149.2024.2423935},
  urldate = {2025-08-17}
}

@article{ryland2025ac,
  title = {Revealing {{Parallel Inter-}} and {{Intra-Ligand Charge Transfer Dynamics}} in [{{Ru}}({{L}})2(Dppz)]2+ {{Molecular Lightswitch}} with {{N K-Edge X-Ray Absorption Spectroscopy}}},
  author = {Ryland, Elizabeth S. and Yang, Xinzheng and Garratt, Douglas and Henke, Wade C. and Kahraman, Abdullah and Taub, Maxwell and Sachs, Michael and Biasin, Elisa and Hampton, Christina Y. and Hoffman, David J. and Coslovich, Giacomo and Kunnus, Kristjan and Dakovski, Georgi L. and Mara, Michael W. and Chen, Lin X. and Mulfort, Karen L. and Li, Xiaosong and Cordones, Amy A.},
  year = 2025,
  journal = {Angewandte Chemie},
  volume = {137},
  number = {36},
  pages = {e202509496},
  issn = {1521-3757},
  doi = {10.1002/ange.202509496},
  urldate = {2025-08-11},
  copyright = {\copyright{} 2025 Wiley-VCH GmbH},
  langid = {english}
}

@article{schlappa2025jsr,
  title = {The {{Heisenberg-RIXS}} Instrument at the {{European XFEL}}},
  author = {Schlappa, Justine and Ghiringhelli, Giacomo and Van Kuiken, Benjamin E. and Teichmann, Martin and Miedema, Piter S. and Delitz, Jan Torben and Gerasimova, Natalia and Molodtsov, Serguei and Adriano, Luigi and Baranasic, Bernard and Broers, Carsten and Carley, Robert and Gessler, Patrick and Ghodrati, Nahid and Hickin, David and Hoang, Le Phuong and Izquierdo, Manuel and Mercadier, Laurent and Mercurio, Giuseppe and Parchenko, Sergii and Stupar, Marijan and Yin, Zhong and Martinelli, Leonardo and Merzoni, Giacomo and Peng, Ying Ying and Reuss, Torben and Sreekantan Nair Lalithambika, Sreeju and Techert, Simone and Laarmann, Tim and Huotari, Simo and Schroeter, Christian and Langer, Burkhard and Giessel, Tatjana and Buchheim, Jana and Gwalt, Grzegorz and Sokolov, Andrey and Siewert, Frank and Buechner, Robby and Vaz Da Cruz, Vinicius and Eckert, Sebastian and Liu, Chun-Yu and Sohrt, Christian and Weniger, Christian and Pietzsch, Annette and Neppl, Stefan and Senf, Friedmar and Scherz, Andreas and F{\"o}hlisch, Alexander},
  year = {2025},
  month = jan,
  journal = {Journal of Synchrotron Radiation},
  volume = {32},
  number = {1},
  pages = {29--45},
  issn = {1600-5775},
  doi = {10.1107/S1600577524010890},
  urldate = {2025-05-06},
  langid = {english}
}

@article{schlotter2012rsi,
  title = {The Soft X-Ray Instrument for Materials Studies at the Linac Coherent Light Source x-Ray Free-Electron Laser},
  author = {Schlotter, W. F. and Turner, J. J. and Rowen, M. and Heimann, P. and Holmes, M. and Krupin, O. and Messerschmidt, M. and Moeller, S. and Krzywinski, J. and Soufli, R. and {Fernndez-Perea}, M. and Kelez, N. and Lee, S. and Coffee, R. and Hays, G. and Beye, M. and Gerken, N. and Sorgenfrei, F. and {Hau-Riege}, S. and Juha, L. and Chalupsky, J. and Hajkova, V. and Mancuso, A. P. and Singer, A. and Yefanov, O. and Vartanyants, I. A. and Cadenazzi, G. and Abbey, B. and Nugent, K. A. and Sinn, H. and Lning, J. and Schaffert, S. and Eisebitt, S. and Lee, W. S. and Scherz, A. and Nilsson, A. R. and Wurth, W.},
  year = {2012},
  month = apr,
  journal = {Review of Scientific Instruments},
  volume = {83},
  number = {4},
  issn = {00346748},
  doi = {10.1063/1.3698294}
}

@article{tumbleson2025sa,
  title = {Thermodynamic Phase Transitions of Nematic Order in Magnetic Helices},
  author = {Tumbleson, Zoey and Morley, Sophie A. and Hollingworth, Emily and Singh, Arnab and Bayaraa, Temuujin and Burdet, Nicolas G. and Saleheen, Ahmad Us and McCarter, Margaret R. and Raftrey, David and Pandolfi, Ronald J. and Esposito, Vincent and Dakovski, Georgi L. and Decker, Franz-Josef and Reid, Alexander H. and Assefa, Tadesse A. and Fischer, Peter and Griffin, Sin{\'e}ad M. and Kevan, Stephen D. and Hellman, Frances and Turner, Joshua J. and Roy, Sujoy},
  year = {2025},
  month = jun,
  journal = {Science Advances},
  volume = {11},
  number = {25},
  pages = {eadt5680},
  publisher = {American Association for the Advancement of Science},
  doi = {10.1126/sciadv.adt5680},
  urldate = {2025-08-17}
}

@article{wernet2015n,
  title = {Orbital-Specific Mapping of the Ligand Exchange Dynamics of {{Fe}}({{CO}})5 in Solution},
  author = {Wernet, Ph and Kunnus, K. and Josefsson, I. and Rajkovic, I. and Quevedo, W. and Beye, M. and Schreck, S. and Gr{\"u}bel, S. and Scholz, M. and Nordlund, D. and Zhang, W. and Hartsock, R. W. and Schlotter, W. F. and Turner, J. J. and Kennedy, B. and Hennies, F. and {de Groot}, F. M. F. and Gaffney, K. J. and Techert, S. and Odelius, M. and F{\"o}hlisch, A.},
  year = {2015},
  month = apr,
  journal = {Nature},
  volume = {520},
  number = {7545},
  pages = {78--81},
  publisher = {Nature Publishing Group},
  issn = {1476-4687},
  doi = {10.1038/nature14296},
  urldate = {2025-07-02},
  copyright = {2015 Springer Nature Limited},
  langid = {english}
}

@article{worner2025nrc,
  title = {Ultrafast Spectroscopy of Liquids Using Extreme-Ultraviolet to Soft-{{X-ray}} Pulses},
  author = {W{\"o}rner, Hans Jakob and Wolf, Jean-Pierre},
  year = {2025},
  month = mar,
  journal = {Nature Reviews Chemistry},
  volume = {9},
  number = {3},
  pages = {185--199},
  publisher = {Nature Publishing Group},
  issn = {2397-3358},
  doi = {10.1038/s41570-025-00692-9},
  urldate = {2025-07-02},
  copyright = {2025 Springer Nature Limited},
  langid = {english}
}

@article{tiedtke2014oea,
  title = {Absolute Pulse Energy Measurements of Soft X-Rays at the {{Linac Coherent Light Source}}},
  author = {Tiedtke, K. and Sorokin, A. A. and Jastrow, U. and Jurani{\'c}, P. and Kreis, S. and Gerken, N. and Richter, M. and Arp, U. and Feng, Y. and Nordlund, D. and Soufli, R. and {Fern{\'a}ndez-Perea}, M. and Juha, L. and Heimann, P. and Nagler, B. and Lee, H. J. and Mack, S. and Cammarata, M. and Krupin, O. and Messerschmidt, M. and Holmes, M. and Rowen, M. and Schlotter, W. and Moeller, S. and Turner, J. J.},
  year = {2014},
  month = sep,
  journal = {Optics Express},
  volume = {22},
  number = {18},
  pages = {21214},
  issn = {1094-4087},
  doi = {10.1364/OE.22.021214},
  urldate = {2025-08-18},
  copyright = {https://doi.org/10.1364/OA\_License\_v1\#VOR-OA},
  langid = {english}
}

@article{larsen2023oe,
  title = {Compact Single-Shot Soft {{X-ray}} Photon Spectrometer for Free-Electron Laser Diagnostics},
  author = {Larsen, Kirk A. and Borne, Kurtis and Obaid, Razib and Kamalov, Andrei and Liu, Yusong and Cheng, Xinxin and James, Justin and Driver, Taran and Li, Kenan and Liu, Yanwei and Sakdinawat, Anne and David, Christian and Wolf, Thomas J. A. and Cryan, James P. and Walter, Peter and Lin, Ming-Fu},
  year = {2023},
  month = oct,
  journal = {Optics Express},
  volume = {31},
  number = {22},
  pages = {35822},
  issn = {1094-4087},
  doi = {10.1364/OE.502105},
  urldate = {2025-08-18},
  langid = {english}
}

@article{sopenamoros2024jacs,
  title = {Tracking {{Cavity Formation}} in {{Electron Solvation}}: {{Insights}} from {{X-ray Spectroscopy}} and {{Theory}}},
  author = {Sopena Moros, Arturo and Li, Shuai and Li, Kai and Doumy, Gilles and Southworth, Stephen H and Otolski, Christopher and Schaller, Richard D and Kumagai, Yoshiaki and Rubensson, Jan-erik and Simon, Marc and Dakovski, Georgi and Kunnus, Kristjan and Robinson, Joseph S and Hampton, Christina Y and Hoffman, David J and Koralek, Jake and Loh, Zhi-heng and Santra, Robin and Inhester, Ludger and Young, Linda},
  year = {2024},
  month = jan,
  journal = {Journal of the American Chemical Society},
  volume = {146},
  number = {5},
  pages = {3262--3269},
  issn = {0002-7863},
  doi = {10.1021/jacs.3c11857}
}

@article{duris2020np,
  title = {Tunable Isolated Attosecond {{X-ray}} Pulses with Gigawatt Peak Power from a Free-Electron Laser},
  author = {Duris, Joseph and Li, Siqi and Driver, Taran and Champenois, Elio G. and MacArthur, James P. and Lutman, Alberto A. and Zhang, Zhen and Rosenberger, Philipp and Aldrich, Jeff W. and Coffee, Ryan and Coslovich, Giacomo and Decker, Franz-Josef and Glownia, James M. and Hartmann, Gregor and Helml, Wolfram and Kamalov, Andrei and Knurr, Jonas and Krzywinski, Jacek and Lin, Ming-Fu and Marangos, Jon P. and Nantel, Megan and Natan, Adi and O'Neal, Jordan T. and Shivaram, Niranjan and Walter, Peter and Wang, Anna Li and Welch, James J. and Wolf, Thomas J. A. and Xu, Joseph Z. and Kling, Matthias F. and Bucksbaum, Philip H. and Zholents, Alexander and Huang, Zhirong and Cryan, James P. and Marinelli, Agostino},
  year = {2020},
  month = jan,
  journal = {Nature Photonics},
  volume = {14},
  number = {1},
  eprint = {1906.10649},
  pages = {30--36},
  publisher = {Springer US},
  issn = {1749-4885},
  doi = {10.1038/s41566-019-0549-5},
  archiveprefix = {arXiv}
}

@article{guo2024np,
  title = {Experimental Demonstration of Attosecond Pump--Probe Spectroscopy with an {{X-ray}} Free-Electron Laser},
  author = {Guo, Zhaoheng and Driver, Taran and Beauvarlet, Sandra and Cesar, David and Duris, Joseph and Franz, Paris L. and Alexander, Oliver and Bohler, Dorian and Bostedt, Christoph and Averbukh, Vitali and Cheng, Xinxin and DiMauro, Louis F. and Doumy, Gilles and Forbes, Ruaridh and Gessner, Oliver and Glownia, James M. and Isele, Erik and Kamalov, Andrei and Larsen, Kirk A. and Li, Siqi and Li, Xiang and Lin, Ming-Fu and McCracken, Gregory A. and Obaid, Razib and O'Neal, Jordan T. and Robles, River R. and Rolles, Daniel and Ruberti, Marco and Rudenko, Artem and Slaughter, Daniel S. and Sudar, Nicholas S. and Thierstein, Emily and Tuthill, Daniel and Ueda, Kiyoshi and Wang, Enliang and Wang, Anna L. and Wang, Jun and Weber, Thorsten and Wolf, Thomas J. A. and Young, Linda and Zhang, Zhen and Bucksbaum, Philip H. and Marangos, Jon P. and Kling, Matthias F. and Huang, Zhirong and Walter, Peter and Inhester, Ludger and Berrah, Nora and Cryan, James P. and Marinelli, Agostino},
  year = {2024},
  month = jul,
  journal = {Nature Photonics},
  volume = {18},
  number = {7},
  pages = {691--697},
  publisher = {Nature Publishing Group},
  issn = {1749-4893},
  doi = {10.1038/s41566-024-01419-w},
  urldate = {2025-08-18},
  copyright = {2024 The Author(s), under exclusive licence to Springer Nature Limited},
  langid = {english}
}

@inproceedings{cocco2013,
  title = {The Optical Design of the Soft X-Ray Self Seeding at {{LCLS}}},
  booktitle = {{{SPIE Optical Engineering}} + {{Applications}}},
  author = {Cocco, Daniele and Abela, Rafael and Amann, John W. and Chow, Ken and Emma, Paul J. and Feng, Yiping and Gassner, Georg L. and Hastings, Jerome and Heimann, Philip and Huang, Zhirong and Loos, Henrik and Montanez, Paul A. and Morton, Daniel and Nuhn, Heinz-Dieter and Ratner, Daniel F. and Rodes, Larry N. and Flechsig, Uwe and Welch, James J. and Wu, Juhao},
  editor = {Klisnick, Annie and Menoni, Carmen S.},
  year = {2013},
  month = sep,
  pages = {88490A},
  address = {San Diego, California, United States},
  doi = {10.1117/12.2024402},
  urldate = {2025-08-18},
  langid = {english}
}

@article{franz2024np,
  title = {Terawatt-Scale Attosecond {{X-ray}} Pulses from a Cascaded Superradiant Free-Electron Laser},
  author = {Franz, Paris and Li, Siqi and Driver, Taran and Robles, River R. and Cesar, David and Isele, Erik and Guo, Zhaoheng and Wang, Jun and Duris, Joseph P. and Larsen, Kirk and Glownia, James M. and Cheng, Xinxin and Hoffmann, Matthias C. and Li, Xiang and Lin, Ming-Fu and Kamalov, Andrei and Obaid, Razib and Summers, Adam and Sudar, Nick and Thierstein, Emily and Zhang, Zhen and Kling, Matthias F. and Huang, Zhirong and Cryan, James P. and Marinelli, Agostino},
  year = {2024},
  month = jul,
  journal = {Nature Photonics},
  volume = {18},
  number = {7},
  pages = {698--703},
  publisher = {Springer US},
  issn = {1749-4885},
  doi = {10.1038/s41566-024-01427-w},
  isbn = {4156602401427}
}

@article{hoffman2025nca,
  title = {Surface Structure of Water from Soft {{X-ray}} Second Harmonic Generation},
  author = {Hoffman, David J. and Devlin, Shane W. and Garratt, Douglas and Jamnuch, Sasawat and Spies, Jacob A. and Nebgen, Bailey R. and Schacher, Daniel and Do, Alexandria and Bernal, Franky and Riffe, Erika J. and Kunnus, Kristjan and Hampton, Christina Y. and Duris, Joseph and Cesar, David and Sudar, Nicholas and Dakovski, Georgi L. and Drisdell, Walter S. and Lawler, Keith V. and Marinelli, Agostino and Zuerch, Michael W. and Saykally, Richard J. and Schwartz, Craig P. and Pascal, Tod A. and Koralek, Jake D.},
  year = 2025,
  month = nov,
  journal = {Nature Communications},
  volume = {16},
  number = {1},
  pages = {10522},
  publisher = {Nature Publishing Group},
  issn = {2041-1723},
  doi = {10.1038/s41467-025-65514-4},
  urldate = {2025-12-16},
  copyright = {2025 The Author(s)},
  langid = {english}
}

@article{jay2018jpcl,
  title = {Disentangling {{Transient Charge Density}} and {{Metal}}--{{Ligand Covalency}} in {{Photoexcited Ferricyanide}} with {{Femtosecond Resonant Inelastic Soft X-ray Scattering}}},
  author = {Jay, Raphael M. and Norell, Jesper and Eckert, Sebastian and Hantschmann, Markus and Beye, Martin and Kennedy, Brian and Quevedo, Wilson and Schlotter, William F. and Dakovski, Georgi L. and Minitti, Michael P. and Hoffmann, Matthias C. and Mitra, Ankush and Moeller, Stefan P. and Nordlund, Dennis and Zhang, Wenkai and Liang, Huiyang W. and Kunnus, Kristjan and Kubi{\v c}ek, Katharina and Techert, Simone A. and Lundberg, Marcus and Wernet, Philippe and Gaffney, Kelly and Odelius, Michael and F{\"o}hlisch, Alexander},
  year = {2018},
  month = jun,
  journal = {The Journal of Physical Chemistry Letters},
  volume = {9},
  number = {12},
  pages = {3538--3543},
  publisher = {American Chemical Society},
  doi = {10.1021/acs.jpclett.8b01429},
  urldate = {2025-08-18}
}

@article{emma2010np,
  title = {First Lasing and Operation of an {\AA}ngstrom-Wavelength Free-Electron Laser},
  author = {Emma, P. and Akre, R. and Arthur, J. and Bionta, R. and Bostedt, C. and Bozek, J. and Brachmann, A. and Bucksbaum, P. and Coffee, R. and Decker, F.-J. and Ding, Y. and Dowell, D. and Edstrom, S. and Fisher, A. and Frisch, J. and Gilevich, S. and Hastings, J. and Hays, G. and Hering, Ph and Huang, Z. and Iverson, R. and Loos, H. and Messerschmidt, M. and Miahnahri, A. and Moeller, S. and Nuhn, H.-D. and Pile, G. and Ratner, D. and Rzepiela, J. and Schultz, D. and Smith, T. and Stefan, P. and Tompkins, H. and Turner, J. and Welch, J. and White, W. and Wu, J. and Yocky, G. and Galayda, J.},
  year = {2010},
  month = sep,
  journal = {Nature Photonics},
  volume = {4},
  number = {9},
  pages = {641--647},
  publisher = {Nature Publishing Group},
  issn = {1749-4893},
  doi = {10.1038/nphoton.2010.176},
  urldate = {2025-08-18},
  copyright = {2010 Springer Nature Limited},
  langid = {english}
}

@book{bergman2017x,
  title={X-ray free electron lasers: applications in materials, chemistry and biology},
  author={Bergman, Uwe and Yachandra, Vittal K and Yano, Junko},
  volume={18},
  year={2017},
  publisher={Royal Society of Chemistry}
}

@article{kahraman2025,
      title={Tracking Electron, Proton, and Solvent Motion in Proton-Coupled Electron Transfer with Ultrafast X-rays}, 
      author={Abdullah Kahraman and Michael Sachs and Soumen Ghosh and Benjamin I. Poulter and Estefanía Sucre-Rosales and Elizabeth S. Ryland and Douglas Garratt and Sumana L. Raj and Natalia Powers-Riggs and Subhradip Kundu and Christina Y. Hampton and David J. Hoffman and Giacomo Coslovich and Georgi L. Dakovski and Patrick L. Kramer and Matthieu Chollet and Roberto A. Mori and Tim B. van Driel and Sang-Jun Lee and Kristjan Kunnus and Amy A. Cordones and Robert W. Schoenlein and Eric Vauthey and Amity Andersen and Niranjan Govind and Christopher Larsen and Elisa Biasin},
      year={2025},
      eprint={2510.03693},
      archivePrefix={arXiv},
      primaryClass={physics.chem-ph},
      url={https://arxiv.org/abs/2510.03693}, 
}

@article{nilsson2013jesrp,
  title = {Resonant Inelastic X-ray Scattering of Liquid Water},
  author = {Nilsson, Anders and Tokushima, Takashi and Horikawa, Yuka and Harada, Yoshihisa and Ljungberg, Mathias P. and Shin, Shik and Pettersson, Lars G. M.},
  year = 2013,
  month = jun,
  journal = {Journal of Electron Spectroscopy and Related Phenomena},
  series = {Progress in {{Resonant Inelastic X-Ray Scattering}}},
  volume = {188},
  pages = {84--100},
  issn = {0368-2048},
  doi = {10.1016/j.elspec.2012.09.011},
  urldate = {2026-02-18}
}

@article{eckert2015apl,
  title = {Principles of Femtosecond {{X-ray}}/Optical Cross-Correlation with {{X-ray}} Induced Transient Optical Reflectivity in Solids},
  author = {Eckert, S. and Beye, M. and Pietzsch, A. and Quevedo, W. and Hantschmann, M. and Ochmann, M. and Ross, M. and Minitti, M. P. and Turner, J. J. and Moeller, S. P. and Schlotter, W. F. and Dakovski, G. L. and Khalil, M. and Huse, N. and F{\"o}hlisch, A.},
  year = 2015,
  month = feb,
  journal = {Applied Physics Letters},
  volume = {106},
  number = {6},
  pages = {061104},
  issn = {0003-6951},
  doi = {10.1063/1.4907949},
  urldate = {2026-03-04}
}

@article{assefa2022rsi,
  title = {The Fluctuation--Dissipation Measurement Instrument at the {{Linac Coherent Light Source}}},
  author = {Assefa, T. A. and Seaberg, M. H. and Reid, A. H. and Shen, L. and Esposito, V. and Dakovski, G. L. and Schlotter, W. and Holladay, B. and Streubel, R. and Montoya, S. A. and Hart, P. and Nakahara, K. and Moeller, S. and Kevan, S. D. and Fischer, P. and Fullerton, E. E. and Colocho, W. and Lutman, A. and Decker, F.-J. and Sinha, S. K. and Roy, S. and Blackburn, E. and Turner, J. J.},
  year = 2022,
  month = aug,
  journal = {Review of Scientific Instruments},
  volume = {93},
  number = {8},
  pages = {083902},
  issn = {0034-6748},
  doi = {10.1063/5.0091297},
  urldate = {2026-03-11}
}

@article{shen2021ma,
  title = {A Snapshot Review---{{Fluctuations}} in Quantum Materials: From Skyrmions to Superconductivity},
  shorttitle = {A Snapshot Review---{{Fluctuations}} in Quantum Materials},
  author = {Shen, L. and Seaberg, M. and Blackburn, E. and Turner, J. J.},
  year = 2021,
  month = may,
  journal = {MRS Advances},
  volume = {6},
  number = {8},
  pages = {221--233},
  issn = {2059-8521},
  doi = {10.1557/s43580-021-00051-y},
  urldate = {2026-03-11},
  langid = {english}
}

@inproceedings{joshi2025,
  title = {Integrated Control Systems for Time-Resolved {{RIXS}} at {{LCLS-II}}: Design and Operational Challenges},
  shorttitle = {Integrated Control Systems for Time-Resolved {{RIXS}} at {{LCLS-II}}},
  booktitle = {20th {{Int}}. {{Conf}}. {{Accel}}. {{Large Exp}}. {{Phys}}. {{Control Syst}}.},
  author = {Joshi, Jyoti and Dakovski, Georgi and Kunnus, Kristjan and Oppermann, Patrick and Yougoubare, Quentin},
  editor = {Wootton, Kent and Biss, Jeffery and Bruno, Gustavo and Kirchman, Jeff and Jaje, Kelly and Skiadopoulos, Denise and Wang, SuYin Grass},
  year = 2025,
  month = nov,
  pages = {1553-1557},
  publisher = {JACoW Publishing},
  issn = {2226-0358},
  doi = {10.18429/JACOW-ICALEPCS2025-THMR013},
  urldate = {2026-03-20},
  copyright = {Creative Commons Attribution 4.0 International},
  isbn = {9783954502554},
  langid = {english}
}

@inproceedings{zamudioestrada2025,
  title = {Liquid Sample Delivery Controls at {{LCLS}}},
  booktitle = {20th {{Int}}. {{Conf}}. {{Accel}}. {{Large Exp}}. {{Phys}}. {{Control Syst}}.},
  author = {Zamudio Estrada, Josue and Hampton, Christina and Cohen, Joshua and Joshi, Jyoti and Sierra, Raymond and Dehe, Sebastian},
  editor = {Wootton, Kent and Biss, Jeffery and Bruno, Gustavo and Kirchman, Jeff and Jaje, Kelly and Skiadopoulos, Denise and Wang, SuYin Grass},
  year = 2025,
  month = nov,
  pages = {1529-1532},
  publisher = {JACoW Publishing},
  issn = {2226-0358},
  doi = {10.18429/JACOW-ICALEPCS2025-THMR004},
  urldate = {2026-03-20},
  copyright = {Creative Commons Attribution 4.0 International},
  isbn = {9783954502554},
  langid = {english}
}

\end{document}